\documentstyle[epsfig,multicol,aps]{revtex}

\newcommand{\infig}[2]{\begin{center}{\epsfig{file=#2,width=#1}}\end{center}} 
\newenvironment{Figure}{\vskip 2pt\noindent}{\vskip 2pt\noindent}
\newcommand{\Caption}[1]{\par\noindent{\small #1} \par}
\newcommand{\Endrule}{\vskip 3pt\noindent\hrule width 8.6cm\vskip 3pt}
\newcommand{\Beginrule}{\vskip 3pt\noindent\hbox{%
\vbox{\hbox to 9cm{\hfill}}\vbox{\hrule width 9cm}} \vskip 3pt}

\begin{document}
\title{Quantum kinetic theory IV: Intensity and amplitude fluctuations of a 
Bose-Einstein condensate at finite temperature including trap loss}

\author{D.~Jaksch$^{1}$, C.W.~Gardiner$^{2}$, K.M. Gheri$^1$ and P.~Zoller$^{1}$}
\address{$^1$ Institut f{\"u}r Theoretische Physik,
Universit{\"a}t Innsbruck, 6020 Innsbruck, Austria}
\address{$^2$ Physics Department, Victoria University, Wellington,
New Zealand}

\maketitle

\begin{abstract} We use the quantum kinetic theory to calculate the
steady state and the fluctuations of a trapped Bose-Einstein condensate
at a finite temperature. The system is divided in a condensate and a
noncondensate part. A quantum-mechanical description based on the
number conserving Bogoliubov method is used for describing the
condensate part. The noncondensed particles are treated as a classical
gas in thermal equilibrium with temperature $T$ and chemical potential
$\mu$. We find a master equation for the reduced density operator of the
Bose-Einstein condensate, calculate the steady state of the system, and
investigate the effect of one-, two-, and three-particle losses on the
condensate. Using linearized Ito equations we find expressions for the
intensity fluctuations and the amplitude fluctuations in the condensate.
A Lorentzian line shape is found for the intensity correlation function
that is characterized by a time constant $\gamma_I^{-1}$ derived in the
paper. For the amplitude correlation function we find ballistic behavior
for time differences smaller than $\gamma_I^{-1}$, and diffusive behavior
for larger time differences. \end{abstract}

%\narrowtext  
\begin{multicols}{2}

% ********************************************************************************
% ***                                                                          ***
% ***                           Introduction                                   ***
% ***                                                                          ***
% ********************************************************************************

\section{Introduction}

In a series of papers we have developed a quantum kinetic (QK) theory with
application to Bose condensation of cold dilute gases. In the first two
papers, 
which we shall refer to as QKI  
\cite{QKI} and QKII \cite{QKII}, we considered
a spatially homogeneous, weakly condensed system, where the interaction
between the atoms was assumed to be sufficiently weak for quasiparticle
effects to be negligible. In 
QKIII \cite{QKIII} the theory was extended
to a strongly condensed gas in a trapping potential under the assumption
that the noncondensed vapor acts as a heat and particle reservoir for
the condensate (see also Ref.~\cite{Anglin}), a situation which corresponds
closely to present experiments of Bose condensation with alkali-metal vapors
\cite{JILA,MIT,RICE,AUSTIN,ROWLAND,STANFORD,Rempe}.

In the present paper (QKIV) we will study the steady state,
amplitude, and phase fluctuations of a trapped Bose-Einstein condensate
at {\em finite temperature}, including the effects of one-, two-, and
three particle losses on the condensate. Such a study seems
particularly timely in view of the present interest in the dynamics and
measurement of the phase of the Bose condensate (for a review see
Ref.~\cite{Villain}). Until now the discussion in the literature has
essentially focused on phase collapse or diffusion, and phase revivals
in the zero-temperature limit, analyzing the dependence of collapse
and revival times on the trap potential, the dimensionality of the gas,
atom number fluctuations, and the coherent dynamics of the
condensate \cite{phase1,phase2,measurementofphase}. In contrast, in the 
present work we will study in detail fluctuations as a result of interaction 
of the condensate with a (reservoir of) uncondensed atoms.

%Start change
We will almost exclusively consider a grand canonical particle reservoir in this work.
This particle reservoir will be assumed to have a constant chemical potential
and not to be influenced by the mean field of the condensate. The results
will therefore only be valid in the case of small condensate and large
thermal particle numbers. In all other cases the simple model used here
for the thermal particles will have to be replaced by a more sophisticated one.
A more detailed discussion on the expected correction to the results
due to particle conservation and mean-field effects is given in Appendix {\ref{cancorr}}.
%End change

The starting point of the theory is to decompose the field operator
describing the $N$-particle system into condensate and
noncondensate parts. The division is based on a splitting into two energy
regions where the high energy band is supposed to contain particles in a
thermal state corresponding to a given temperature $T$, and chemical
potential $\mu$. The condensate band contains the actual condensate as
well as quasi-particle excitations. A quantum-mechanical description
based on the number conserving Bogoliubov method is used for describing
the condensate part \cite{NumCon}. To facilitate the analysis we
drop the quasi-particles and describe the condensate by a single
mode with destruction operator $B$, and the spatial wave function
$\xi_N({\bf x})$, corresponding to the solution of the Gross-Pitaevskii
equation for $N$ condensate particles. Elimination of the noncondensate
part leaves us with a master equation for the
(reduced) condensate density operator. The physics contained in this
equation is quite rich. The master equation accounts for loss/gain of
particles to/from the thermal band, and
phase-destroying but number-conserving collisions between condensate 
and noncondensate particles as well as linear and nonlinear trap loss. 

A diagrammatic illustration of the processes described by the master
equation is given in Fig.~1 %\ref{processes}
. We realize that there are
two types of processes. On the one hand, there are those which involve
particles from both the condensate and the noncondensed band.
They comprise processes which lead to a redistribution of particles
between the two bands at rates $W^+(N)$ (condensate gain) and $W^-(N)$
(condensate loss), as well as number conserving scattering events of
thermal particles off the condensate. The latter occur with a rate
$R_{00}(N)$ and will give rise to fluctuations in the condensate phase.
Explicit expressions for $W^{\pm}$ and $R_{00}$ will be given below. On
the other hand there are several loss mechanisms at work which deplete
the condensate \cite{JilaCCC,Kettrev}. There is one-particle loss due 
to collisions with
background gas atoms with associated rate $\gamma_1$. Two particles can
be lost with rate $\gamma_2(N)$ from the condensate if two condensed
particles collide, and one of them changes its internal state. This
particle no longer sees the trap, and escapes. Its partner is imparted with
the energy difference set free by the collision, and is also lost from
the trap. Finally, three-particle loss can occur with rate $\gamma_3(N)$
if in a three-particle collision a dimer is formed. The binding energy
is imparted to the third particle, and all of them escape from the trap.
Note that the description of the noncondensate particles in terms of a
thermal reservoir results in a finite occupation of the condensate mode
even in the presence of loss channels. 

\begin{Figure}
\infig{5cm}{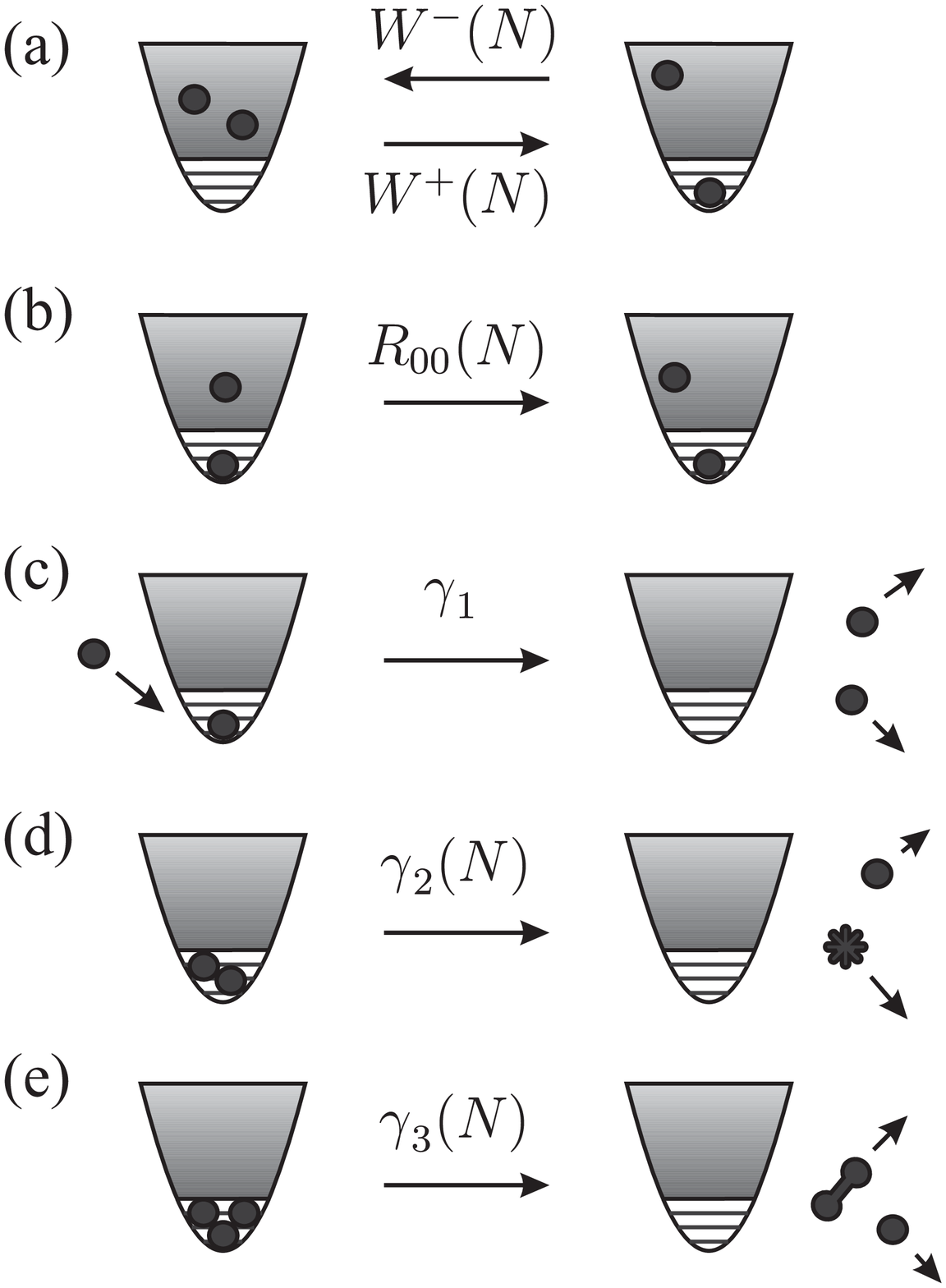}
\Caption{Fig.~1. Interpretation of the
processes described by the master equation: (a) $W^+(N)$ and $W^-(N)$
are the feeding and depletion rates of the condensate from and to the
noncondensed thermal band of particles, respectively. (b) $R_{00}$ is
the rate of thermal particles bouncing off the condensate without
changing its occupation number. $\gamma_1$ (c), $\gamma_2(N)$ (d)
and $\gamma_3(N)$ (e) are the rates of one,- two,- and three-particle
losses, respectively. We use a star to indicate a change in the internal
state of a particle. The barbell represents two atoms having formed a
molecule.\label{processes}}
\end{Figure}

% ********************************************************************************
% ***                                                                          ***
% ***                           Main results                                   ***
% ***                                                                          ***
% ********************************************************************************

\section{Main results}

In this section we will give a short overview of the main results to be
derived in detail in the remainder of this paper. As a starting point
for our analysis of the particle number and phase fluctuations of a
Bose-Einstein condensate at a finite temperature, we adopt the
theoretical description developed in the precursor papers QKI-III
 \cite{QKI,QKII,QKIII}. 

\subsection{Fluctuation analysis}

In the limit of large condensate particle numbers, we may approximate 
the master equation [Eq.~(\ref{master})]
by an equation that is of Lindblad type. This has the advantage that
standard techniques developed in quantum optics for the description of
fluctuation properties become applicable \cite{QNoise}. We have thus derived quantum
stochastic differential equations for the condensate particle number
$\tilde{N}=B^\dagger B$ and the Glogower-Susskind phase operator
$e^{i\phi}$, which is known to characterize phase fluctuations well in
the limit of large occupation numbers \cite{KGLaser}. Linearization of these equations
is permissible in the very same limit of large average occupation
$\bar{N}=\langle \tilde{N} \rangle$, and allows us to work out the
two-time correlation functions of occupation number and phase. The
spectra of condensate occupation number and amplitude fluctuations are
then immediately obtainable by Fourier transformation.

\subsubsection{Condensate particle number fluctuations}

For the correlation function of the particle number fluctuations \cite{corrinout}
we obtain the following result which
holds in the stationary limit, i.e., for times satisfying $t+s \gg \gamma_I^{-1}$
\begin{equation}
\langle \tilde{N}(t), \tilde{N}(s) \rangle=\frac{\bar{f}_2}{2 \gamma_I} 
\mbox{e}^{-\gamma_I \left|t-s\right|}=\sigma^2\mbox{e}^{-\gamma_I 
\left|t-s\right|},\label{eq:intcorr}
\label{intf}
\end{equation}
with $\langle a,b \rangle= \langle a b \rangle- \langle 
a \rangle \langle b \rangle$.
In Eq.~(\ref{intf}) 
 \begin{equation}
 \bar{f}_2/2=\bar{N}( \bar{W}^+ +\bar{W}^- +\gamma_1)+4 \bar{N}^2 \bar{\gamma}_2 + 
9\bar{N}^3 \bar{\gamma}_3,
 \end{equation}
with a bar denoting evaluation at $N=\bar{N}$. 
Note that $\sigma$ appearing in Eq.~(\ref{eq:intcorr}) is a measure of the width of the 
particle number distribution.
The characteristic time constant $\gamma_I$ with which the 
particle number fluctuations regress is given by
\begin{equation}
\gamma_I=2\bar{N}\partial_{\bar{N}}\bar{W}^- 
-(\bar{\gamma}_2+3\bar{\gamma}_3\bar{N})8\bar{N}/5.
\label{gammaIeq}
\end{equation}
%Start change
$\partial_{\bar{N}}$ denotes the derivative with respect to $N$ and
evaluation at $N=\bar{N}$.
%End change
The exact size of this rate depends on the specific experimental setup.
A convenient quantity to assess the amount of fluctuations present 
is the well known Mandel $Q$ parameter defined as 
\begin{equation}
Q = \lim_{t \rightarrow \infty}\langle \tilde{N}(t), \tilde{N}(t) \rangle/\langle \tilde{N}(t)\rangle -1.
\end{equation}
A coherent state would correspond to a value of $Q=0$, while a number state yields the minimum 
value $Q=-1$.
Assuming two and three particle losses to be insignificant we find for the Mandel $Q$ parameter,
\begin{equation}
Q \approx \frac{ 5 kT}{2 \mu_{\bar{N}}}-1.
\end{equation}
To arrive at this result the approximate expressions of the rates $W^+$ and $W^-$ as given in 
Eqs.~(\ref{eq:Wpm}) in terms
of the scattering length $a$, the reservoir temperature $T$ and chemical potential $\mu$ have 
been used.

The results (for details, see Sec.~\ref{corrfct}) can now be summarized as follows. 
The variance of the occupation of the condensate
mode is proportional to $kT/\mu_{\bar{N}}$, with $\mu_{\bar{N}}$ the chemical potential
for the condensate mode with average occupation $\bar{N}$. In the Thomas-Fermi approximation
$\mu_{N}$ is given by
\begin{equation}
\mu_N=\left(\frac{15 N u m^{3/2} \omega_x \omega_y \omega_z}{16 \sqrt{2} \pi}\right)^{2/5}.
\label{ThFchem}
\end{equation}
The constants used in Eq.~(\ref{ThFchem}) are defined in Sec.~\ref{dessys}.
The characteristic rate at which the 
correlation function drops off is roughly given by $\gamma_I\approx 2\bar{N}
\partial_{\bar{N}} \bar{W}^-$, with $W^-$ the loss rate to the noncondensed band. 

\subsubsection{Amplitude and phase fluctuations}

In the limit of large and well-defined average occupation number $\bar{N}$ of the condensatemmode,
the amplitude correlation function $\langle B^\dagger(t)B(s)\rangle $ is well suited to
assess phase properties of the condensate mode \cite{corrinout}. In particular, the spectrum of 
phase fluctuations
is identical to the  spectrum of amplitude fluctuations.
For the amplitude correlation function we obtain for $t>s$
\begin{equation}
\langle B^{\dagger}(t) B(s) \rangle = {\bar{N}} \mbox{e}^{\left( i\frac{\mu_{\bar{N}}}{\hbar}
 - \frac{16}{25} \bar{R}_{00} \right) \left( t-s \right)
-\eta \left( \gamma_I (t-s)+
\mbox{e}^{-\gamma_I (t-s)} -1\right)},
\label{cfct}
\end{equation}
where $\eta=\left({\sigma \partial_{\bar{N}}\mu_{\bar{N}}}/{\gamma_I \hbar}\right)^2$.
As we will see in Sec.~\ref{numv} $R_{00}$ is negligible in the
above correlation function [Eq.~(\ref{cfct})] for most of the current experiments.
The structure of the correlation function indicates that there are two distinct
time regimes:
\begin{itemize}
\item For $\gamma_I|t-s|\ll1$, 
the term proportional to $\eta$ in the exponent is proportional to
$(t-s)^2$. This is called the ballistic regime.
\item For $|t-s|\gamma_I\ge 1$, the phase behaves like that of a system undergoing phase diffusion.
A characteristic of such behavior is a linear dependence of the exponent on $|t-s|$.
Note that for large time differences we observe the legacy of the ballistic regime
in the form of a rescaling of $\bar{N}$ to $\bar{N}_\infty=\bar{N}e^{\eta}$.    
\end{itemize} 

\subsubsection{Numerical values}
\label{numv}

Using data from the experiments at JILA \cite{JilaExp} and an average occupation number of 
$\bar{N}=25000$ Rubidium atoms at a temperature $T=0.5 \mu K$ in a trap with
$f_x=f_y=f_z/\sqrt{8}=47 {\rm Hz}$ we obtain for the rates $\gamma_I \approx 2 {\rm Hz}$,
$\bar{R}_{00} \approx 4 {\rm mHz}$, $\bar{W}^{\pm}\approx 50 {\rm Hz} $, and $\eta \approx 
800$.
These values have to be understood as order of magnitude estimates for current
experiments. Note that we used a value of $a=2.6 {\rm nm}$ for the scattering letngth of
rubidium whereas recent experiments at JILA showed that $a=5.1 {\rm nm}$.

% ********************************************************************************
% ***                                                                          ***
% ***                               Model                                      ***
% ***                                                                          ***
% ********************************************************************************

\section{Model}

In this section we briefly describe the basic concepts of the quantum kinetic theory
developed in Ref.~\cite{QKI,QKII,QKIII,NumCon,BosGro}. We do not give a detailed derivation
of the master equation used throughout the paper since this can be found in Ref.~\cite{QKIII}.

\subsection{Description of the system}
\label{dessys}

The Hamiltonian of a weakly interacting Bose gas confined by a potential
$ V_T({\bf x}) $ in second quantization is written as
\begin{eqnarray}
H&=&\int d^3{\bf x} \, \psi^{\dagger}({\bf x})
\left(- {\hbar ^2 \over 2m} \nabla^2 + V_T({\bf x}) \right) \psi ({\bf x})   
\nonumber \\ &&
+{1\over 2}\int d^3{\bf x}\int d^3{\bf x}' \, \psi ^{\dagger }({\bf x})
\psi ^{\dagger }( {\bf x}') u( {\bf x}-{\bf x}') 
\psi ( {\bf x}') \psi ( {\bf x}).
\label{Hamiltonian}
\end{eqnarray}
$\psi ({\bf x})$ is the standard bosonic field operator.
The two-body interaction potential $u({\bf x}-{\bf x}')$ is a short-range potential of the form $u \delta({\bf x}-{\bf x}')$ where $u=4 \pi \hbar a /m$ 
with $a$ the scattering length. We assume the trapping potential to be of the form 
$V_T({\bf x})=m(\omega_x^2 x^2 + \omega_x^2 y^2 + \omega_z^2 z^2)/2$.

As was shown in Ref.~\cite{QKIII} we can obtain a master equation for the density
operator of the condensate mode by dividing the Bose gas into two energy regions 
called the condensate band $R_C$ and the noncondensate band $R_{\rm NC}$. The boundary $E_R$ 
between these two regions is chosen according to Ref.~\cite{QKIII} such that 
$R_{\rm NC}$ is not significantly affected by the mean field of the condensate.
The field operator is then written as $\psi ({\bf x})=\psi_{\rm NC} ({\bf x})+
\phi ({\bf x})$ describing particles in the noncondensate band and in the 
condensate band, respectively.
We will treat the particles in $R_{\rm NC}$ as classical thermalized particles 
characterized by a temperature $T$ and a chemical potential $\mu$.
The particles in $R_C$ are affected by the mean field, and they have to be treated 
quantum-mechanically. $R_C$ contains all the trap levels that are significantly modified by
the mean field of the condensate. As in Ref.~\cite{BosGro} we will use the simplest 
possible description of the condensate band by assuming that only one 
mode, namely, the condensate mode itself,
is important and all the other modes are negligible. The master
equation we will use for our calculations is an equation for the reduced density operator 
$\rho_c$
of the condensate band $R_C$ interacting with the bath of particles in $R_{\rm NC}$. 

\subsubsection{Condensate band $R_C$}

We use the number-conserving Bogoliubov method derived in 
Ref.~\cite{NumCon} to describe the particles in $R_C$. In this formulation
we can write down the condensate band field operator $\phi ({\bf x})$ 
in the form
\begin{eqnarray}
\phi({\bf x)} &=& B(N)\left\{\xi_N({\bf x}) 
+\sum_m{ b_m f_m({\bf x}) + b_m^\dagger g_m({\bf x})\over\sqrt{N}}
\right\}.
\label{quasi}
\end{eqnarray}
The annihilation operator $B(N-1)$ brings the 
system $R_C$ from the ground state with $N$ atoms to the ground state with $N-1$ atoms. 
The action of the operator $B(N-1)$
to the condensate is depicted in Fig.~2 %\ref{Baction}
. 
$|N\rangle_N$ denotes the ground state with $N$ particles in the condensate. 
Applying the operator $B$ to this state yields $B |N\rangle_N=\sqrt{N}|N-1\rangle_{N-1}$.
Note that the operator $B$ changes the chemical potential of the condensate from
$\mu_N$ to $\mu_{N-1}$. 
\begin{Figure}
\infig{5cm}{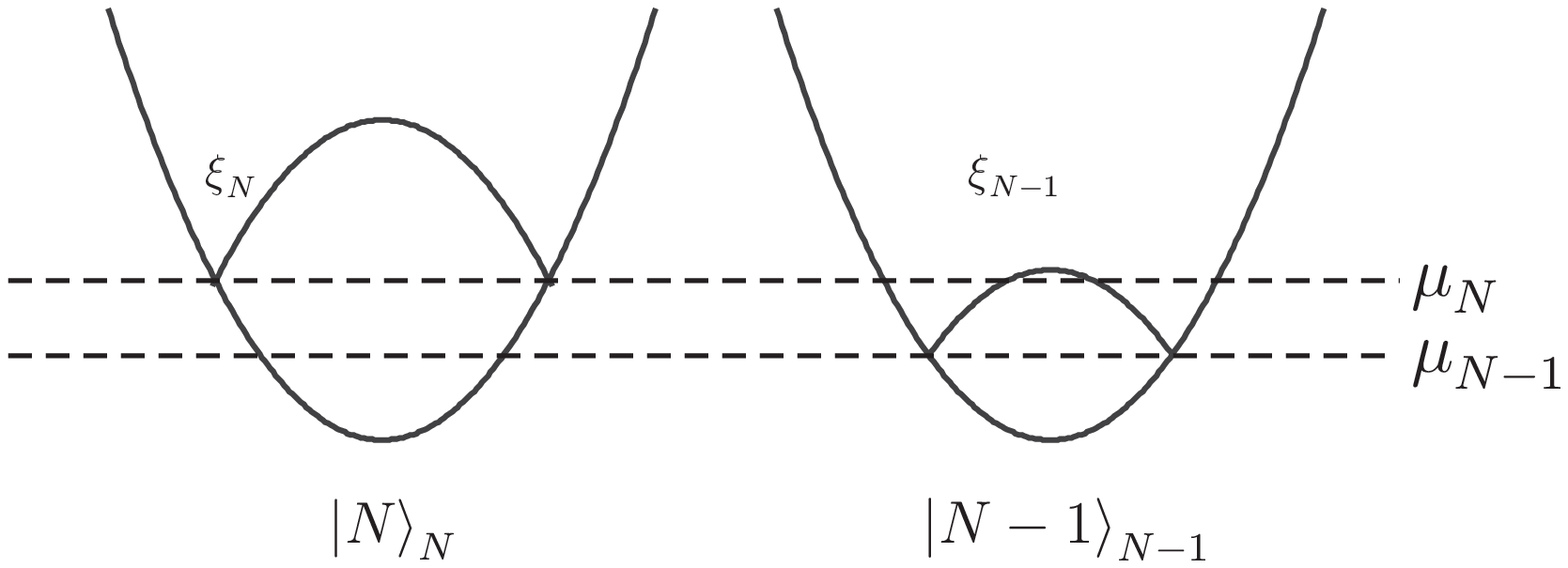}
\Caption{Fig.~2. Condensate ground state with $N$ particles $|N\rangle_N$ and with $N-1$ particles 
$|N-1\rangle_{N-1}$. These two states are connected by the operator $B$, as described in the 
text.
\label{Baction}}
\end{Figure}

As shown in Ref.~\cite{NumCon} $\xi_N({\bf x})$ is the condensate wave function, and satisfies 
the Gross-Pitaevskii equation
\begin{eqnarray}
-{\hbar^2\over 2m}\nabla^2\xi_N 
+V_T\xi_N + N u \bigl |\xi_N\bigr |^2 \xi_N &=& \mu_N \xi_N.
\label{GPE}
\end{eqnarray}
In all our calculations we will use the Thomas-Fermi approximation for the chemical
potential [Eq.~(\ref{ThFchem})] and for the condensate wave function 
\begin{equation}
\xi_N({\bf x})=\sqrt{\frac{\mu_N-V_T({\bf x})}{N u}} \quad
{\text{ for }} \mu_N>V_T({\bf x}),
\label{condxi}
\end{equation}
and zero elsewhere.
%Start change
The energy of the condensate is given by
\begin{equation}
E_0=\frac{5 N \mu_N}{7}
\label{condE0}
\end{equation}
%End change

The amplitudes $f_m({\bf x})$ and $g_m({\bf x})$ describe creation and
destruction of quasiparticles. They are defined in Ref.~\cite{QKIII},
but we will not need them in this paper. We will assume there is always a
fairly large mean number of particles $\bar{N}$ in the condensate so
that we can neglect the influence of the quasiparticles on $\phi({\bf x})$
%Start change
\cite{quasinegl}.
%End change

\subsubsection{Noncondensate band $R_{NC}$}

We treat the noncondensate band as a thermal bath of particles in thermal equilibrium.
According to Ref.~\cite{QKIII} we only need the phase-space density of the noncondensed
particles $F({\bf K},{\bf x})$ to calculate all the rates appearing in the master equation 
for $\rho_c$ [Eq.~(\ref{master})]. In our calculations we will use the classical approximation
\begin{equation}
F({\bf K},{\bf x})=\mbox{e}^{(\mu-\frac{\hbar^2 {\bf K}^2}{2m}-V_T({\bf x}))/kT}.
\label{thermaldist}
\end{equation}
We expect corrections to the rates $W^{\pm}$ and $R_{00}$ from using a more detailed
description of the noncondensed particles. However, all the calculations presented 
here will remain valid, since they are mostly independent of the functional form of 
the rates $W^{\pm}$ and $R_{00}$.
%Start change
We only require that it is permissible to linearize the rates $W^{\pm}(N)$ and $R_{00}(N)$ 
around the mean number of particles in the condensate $\bar{N}$. In the present work we will
mainly present results obtained by using Eqs.~(\ref{eq:Wpm}) and (\ref{R00}) for the
rates $W^{\pm}(N)$ and $R_{00}(N)$ which were obtained by the assumption of a grand 
canonical classical bath of particles not influenced by mean-field effects. Corrections 
to the stationary state due to conservation of the total number of particles as 
well as mean-field effects and quantum statistics effects are estimated in 
Appendix \ref{cancorr}. A further, more detailed, discussion of different models for 
the uncondensed bath (i.e., the noncondensate band) 
lies beyond the scope of this paper, and will be presented in other 
work. However, we want to point out that these models might include canonical baths with 
a constant particle number \cite{QKV}, evaporatively cooled baths \cite{QKVI} as well as 
baths that are continuously fed with particles \cite{AtomLaser}.
%End change

\subsubsection{Trap loss}

There are several processes leading to losses of condensate particles from the trap.
We will consider one,- two,- and three-particle loss with loss rates $\gamma_1$,
$\gamma_2(N)$ and $\gamma_3(N)$, respectively. One-particle loss might be caused by 
background gas particles hitting condensate particles or by coupling condensate 
particles out of the trap as described in Ref.~\cite{corrinout}.
Inelastic two-particle collisions changing the internal properties of the particles
lead to two particle loss. In most of the current experiments the two particle loss
is negligible compared to the three particle loss caused by the inelastic collision of
three particles. In the Thomas-Fermi approximation we obtain for the loss rates
\begin{equation}
\gamma_1 = \frac{K_1}{2} \int d^3x \, \left|\xi_N (x)\right|^2= \frac{K_1}{2}, 
\end{equation}
\begin{eqnarray}
\gamma_2(N) &=& \frac{K_2}{4} \int d^3x \, \left|\xi_N (x)\right|^4= \frac{K_2 16 \cdot  \pi 
\mu_N^{7/2} \sqrt{2}}{105 m^{3/2} \omega_x \omega_y
\omega_z N^2 u^2} \nonumber \\
& \propto & N^{-3/5},
\end{eqnarray}
\begin{equation}
\gamma_3(N) = \frac{K_3}{6} \int d^3x \, \left|\xi_N (x)\right|^6= 
\frac{4 K_3 \mu_{N} \gamma_2(N)}{9 K_2 N u}  \propto N^{-6/5}. 
\end{equation}
For Rubidium the constants $K_i$ have been measured in \cite{JilaCCC}.
The rates $K_i$ have been calculated analytically in 
Refs.~\cite{Anarates1,Anarates2,Anarates3}.

\subsection{Master equation}

% QKIIIEqNummer
We simplify Eq.~(50) in Ref.~\cite{QKIII} for the case that the condensate band consists
only of the condensate mode alone. In contrast to Ref.~\cite{QKIII} we keep 
terms including $R_{00}$, and add the
loss terms to the master equation (see Fig.~1 %
). We thus obtain the following 
equation
for the reduced density operator of the condensate band $\rho_c$:
\end{multicols}
\Endrule
%\widetext
\begin{mathletters}\label{master}
\begin{eqnarray}
\dot{\rho_c} &=& -\frac{i}{\hbar} \left[H_0,\rho_c \right]   \\ 
&+&2 B^{\dagger} \{W^+(\hat{N}) \rho_c \} B - [B B^{\dagger},\{W^+(\hat{N}) \rho_c \}]_+ + 
2 B \{W^-(\hat{N}) \rho_c \} B^{\dagger} - [B^{\dagger} B,\{W^-(\hat{N}) \rho_c \}]_+ 
\label{masterdepl} \\
&+&2 B B^{\dagger} \{R_{00}(\hat{N}) \rho_c \} B B^{\dagger} - [B B^{\dagger} B
B^{\dagger},\{R_{00}(\hat{N}) \rho_c \}]_+   \\
&+&2 B \{\gamma_1 \rho_c \} B^{\dagger} - [B^{\dagger} B,\{\gamma_1 \rho_c \}]_+ +
2 BB \{\gamma_2(\hat{N}) \rho_c \} B^{\dagger} B^{\dagger} - \left[
B^{\dagger} B^{\dagger} BB, \{\gamma_2(\hat{N}) \rho_c \} \right]_+ \label{12loss} \\
&+&2 BBB \{\gamma_3(\hat{N}) \rho_c \} B^{\dagger} B^{\dagger} B^{\dagger} - \left[
B^{\dagger} B^{\dagger} B^{\dagger} BBB, \{\gamma_3(\hat{N}) \rho_c \} \right]_+  
\label{3loss}, 
\end{eqnarray}
\end{mathletters}
%\narrowtext
\Beginrule
\begin{multicols}{2}\noindent 
where $\hat{N} \rho_c =\left[B^{\dagger}B,\rho_c\right]_+/2$.
An intuitive interpretation of the processes described by the master equation 
[Eq.~(\ref{master})]
is given in Fig.~1 %\ref{processes}
.
The free evolution of the condensate is described by the Hamiltonian $H_0$,
\begin{equation}
H_0=E_0(\tilde{N}) \quad \mbox{where} \quad E_0(N+1)-E_0(N)=\mu_N,
\end{equation}
and $\mu_N$ is the chemical potential obtained from the Gross-Pitaevskii equation.

Growth and depletion of the condensate due to interaction with the noncondensed 
particles is given in Eq.~(\ref{masterdepl}).
The growth and the depletion rates $W^{\pm}(N)$ of the condensate were already
calculated in Ref.~\cite{QKIII} using the Maxwell-Boltzmann approximation for the
phase-space density of the noncondensed particles. They are given by 
\begin{mathletters}\label{eq:Wpm}
\begin{equation}
W^+(N)=\frac{4 m (a k T)^2}{ \pi \hbar^3} \mbox{e}^{2 \mu/kT} \left\{\frac{\mu_N}{kT}
K_1(\frac{\mu_N}{kT})\right\},
\end{equation}
\begin{equation}
W^-(N)=\mbox{e}^{(\mu_N-\mu)/kT} W^+(N).
\end{equation}
\end{mathletters}
%

% QKIIIEqNummer
The rate $R_{00}$ defined in Eq.~(141) of Ref.~\cite{QKIII} can be understood as describing the
process of thermal particles bouncing off the condensate. This process does not change the
number of particles in the condensate, but does cause phase fluctuations.
Using the above expression [Eq.~(\ref{thermaldist})] for $F({\bf K},{\bf x})$ and the 
Thomas-Fermi 
approximation for the condensate wave function, we find 
\begin{mathletters}\label{R00}
\begin{equation}
R_{00}(N)=\frac{4 k T \mu_N^4}{9 \pi^4 \hbar^5 \omega_x^3 \omega_z N^2} \mbox{e}^{\mu/kT}
\frac{\mbox{arsinh}\left(\sqrt{\frac{\omega_z^2-\omega_x^2}{\omega_x^2}}\right)}
{\sqrt{\frac{\omega_z^2-\omega_x^2}{\omega_x^2}}}
\label{R001}
\end{equation}
for $\omega_z>\omega_x=\omega_y$, and
\begin{equation}
R_{00}(N)=\frac{4 k T \mu_N^4}{9 \pi^4 \hbar^5 \omega_x^3 \omega_z N^2} \mbox{e}^{\mu/kT}
\frac{\arcsin\left(\sqrt{\frac{\omega_x^2-\omega_z^2}{\omega_x^2}}\right)}
{\sqrt{\frac{\omega_x^2-\omega_z^2}{\omega_x^2}}}
\label{R002}
\end{equation}
\end{mathletters}
for $\omega_z<\omega_x=\omega_y$. A detailed derivation of $R_{00}$ is given in Appendix \ref{R00cal}.

Trap loss is accounted for by the last two lines of the master equation 
Eqs.~(\ref{12loss}) and (\ref{3loss}).
Note that the only difference between the process including the rate
$W^-(N)$ and the one-particle loss rate $\gamma_1$ is the dependence of the two rates on 
the properties $\mu$ and $T$ of the noncondensate.

% ********************************************************************************
% ***                                                                          ***
% ***                      Solutions of the master equation                    ***
% ***                                                                          ***
% ********************************************************************************

\section{Solutions of the master equation}

In this section we investigate solutions of Eq.~(\ref{master}). We find the
stationary solution and derive a differential equation for the mean
number of particles in the condensate $\bar{N}$. Using linearized Ito equations, we
obtain the correlation functions $\langle N(t), N(s) \rangle$ and $\langle B^{\dagger}
(t) B(s) \rangle$. 

We define
\begin{equation}
g_N^{\nu}:= {}_{N}\langle N-\nu|\rho_c|N\rangle_N.
\end{equation}
From Eq.~(\ref{master}), we derive the evolution equation of the matrix elements,
expand the terms in this equation for large $N$ in a Kramers-Moyal
type expansion to order $1/N$, and obtain 
\end{multicols}
\Endrule
%\widetext
\begin{eqnarray}
\dot{g}_N^{\nu} &=& 2 \sqrt{(N-\nu)N} \left\{W^+(N-1) g_{N-1}
^{\nu} - W^-(N) g_N^{\nu} \right\}  \nonumber \\
&+&2 \sqrt{(N-\nu+1)(N+1)} \left\{W^-(N+1) g_{N+1}
^{\nu} - W^+(N) g_{N}^{\nu}\right\}  \nonumber \\
&+& 2 \sqrt{(N-\nu+1)(N+1)} \gamma_1 g_{N+1}^{\nu}-(2N-\nu) \gamma_1 g_N^{\nu} \nonumber \\
&+& 2 \sqrt{(N-\nu+1) (N-\nu+2) (N+1) (N+2)} \gamma_2(N) g_{N+2}^{\nu}-((N-\nu)(N-\nu-1) 
+N(N-1)) \gamma_2(N) g_N^{\nu}  \nonumber \\
&+& 2 \sqrt{(N-\nu+1)(N-\nu+2) (N-\nu+3) (N+1) (N+2) (N+3)} \gamma_3(N) g_{N+3}^{\nu}
\nonumber \\
&-& ((N-\nu)(N-\nu-1)(N-\nu-2)+N(N-1)(N-2)) \gamma_3(N) g_N^{\nu} - (1/\tau_N^{\nu}) 
g_N^{\nu}.
\label{matrixeq}
\end{eqnarray}
The time constant
\begin{equation}
\tau_N^{\nu}=\left\{\frac{\nu^2}{4N}\left[W^+(N)+W^-(N) \right] + \nu^2
R_{00}(N) - \frac{i \nu \mu_N}{\hbar} \right\}^{-1}.
\end{equation}
%
%\narrowtext
\Beginrule
\begin{multicols}{2}
\noindent
determines the time scale on which the non-diagonal matrix elements decay.
We have assumed that $\nu \ll \bar{N}$, and therefore approximated
\begin{equation}
\frac{i}{\hbar}\left(E_0(N)-E_0(N-\nu)\right)\approx \frac{i \nu \mu_N}{\hbar}.
\end{equation}

\subsection{Stationary solution}
\label{statsolsec}

First we neglect trap loss and solve Eq.~(\ref{matrixeq}) for the 
stationary case. All $g_{N}^{\nu}$ with $\nu \neq 0$ drop off exponentially 
in time, and we are left with the diagonal terms only.
We use the detailed balance condition to calculate the stationary
solution $\bar{g}_{N}^{0}$:
\begin{equation}
\frac{\bar{g}_{N+1}^{0}}{\bar{g}_{N}^{0}}=\frac{W^+(N)}{W^-(N+1)} \approx \mbox{e}^
{(\mu-\mu_N)/kT},
\end{equation}
and therefore
\begin{equation}
\bar{g}_{N}^{0} \propto \prod_{n=0}^N \mbox{e}^{(\mu-\mu_n)/kT}.
\end{equation}
Using the Thomas-Fermi approximation for the chemical potential of the condensate,
and replacing the sum occurring in the exponent by an integral, we obtain
\begin{equation}
\bar{g}_{N}^{0} \propto \mbox{e}^{N ( \mu-5 \mu_N /7)/kT}.
\label{exdist}
\end{equation}
This particle distribution has one maximum determined by the condition $\mu_N=\mu$, as expected
from thermodynamics. The position of the maximum differs from the mean number of
particles in the condensate by an amount of order $1/\bar{N}$. 

%Start change
\subsubsection{Linearized solution}

In the case of $\bar{N} \gg 1$ we may linearize the solution Eq.~(\ref{exdist}) around the
mean number of particles in the condensate $\bar{N}$. We
approximate the distribution in Eq.~(\ref{exdist}) by a Gaussian
\begin{equation}
\bar{g}_N^{0} \approx \frac{1}{\sqrt{2 \pi} \sigma} \mbox{e}^{-\frac{(N-\bar{N})^2}{2 
\sigma^2}}.
\label{Gaussappr}
\end{equation}
The width $\sigma$ of this Gaussian is given by
\begin{equation}
\sigma=\sqrt{\frac{5 k T \bar{N}}{2 \mu_{\bar{N}}}}.
\end{equation}
In the Thomas-Fermi approximation $\sigma$ scales with the mean number
of condensate particles like $\bar{N}^{3/10}$. The difference between the
linearized and nonlinearized solutions can therefore only be seen for
very small condensates. Figure 3 %\ref{statsolfig}
 shows a comparison between the Gaussian 
approximation and numerical solutions obtained from Eqs.~(\ref{master}) and (\ref{FPEloss}).
As expected both solutions agree very well with each other even for a mean occupation
of the condensate of only about $\bar{N} \approx 500$.
Note also that the same result for the variance may be obtained from statistical mechanics.
Thus the restriction $\mu_{\bar{N}}/kT > 1$ is not necessary for this result to be valid. 
%End change
\begin{Figure}
\infig{5cm}{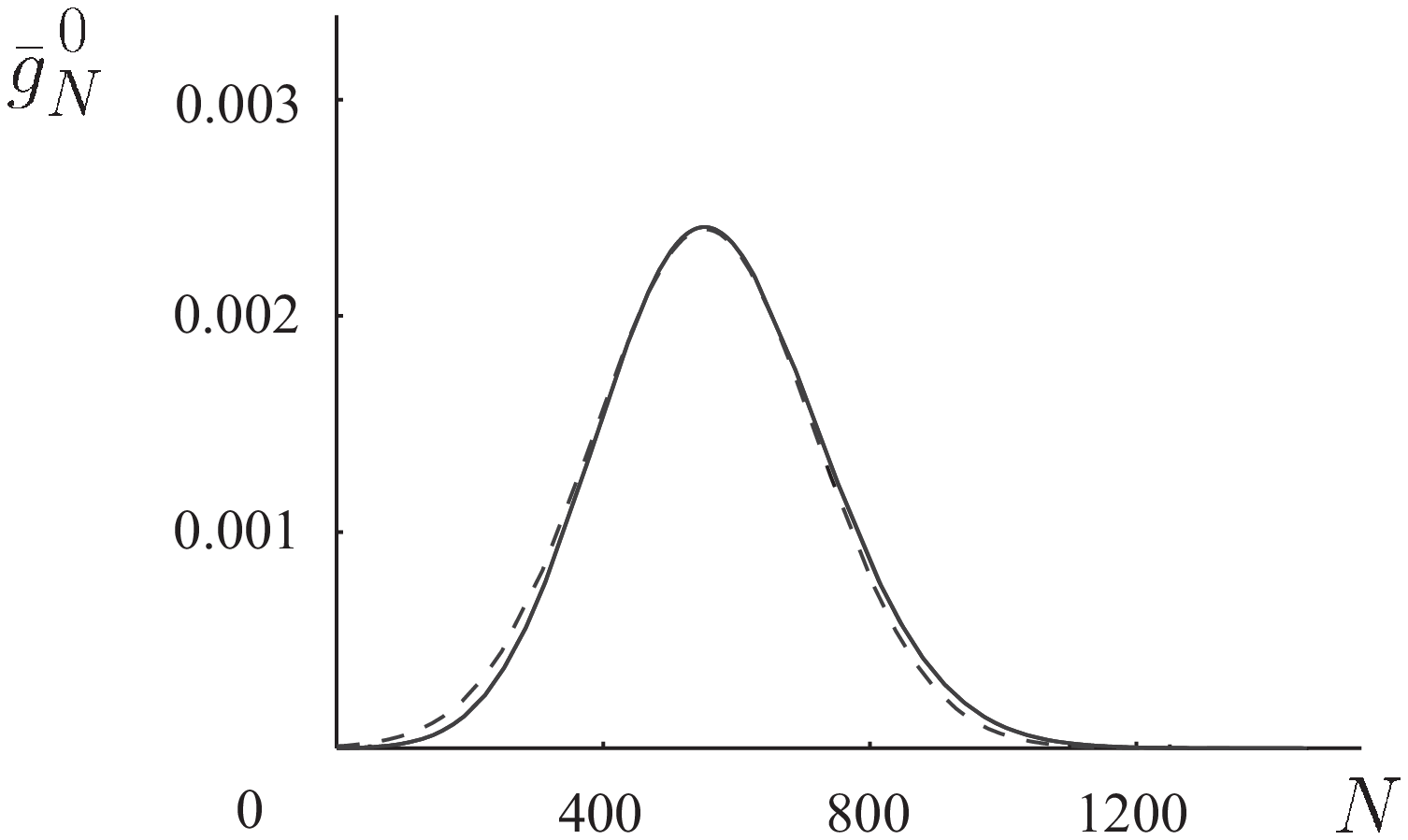}
\Caption{Fig.~3. Stationary particle distribution in the condensate. The trap loss is assumed to be
zero. $\mu/kT=0.05$ and $\mu_{N=1}/kT=0.004$. The solid line represents the numerical 
stationary solution of the Fokker-Planck equation [Eq.~(\ref{FPEloss})] and the detailed balance 
solution of the master equation [Eq.~(\ref{master})]. The dashed line depicts the Gaussian 
approximation [Eq.~(\ref{Gaussappr})]. The trap frequencies are chosen to be 
$f_x=f_y=f_z/{\protect \sqrt{8}}=47 {\rm Hz}$.  The calculation is done for Rubidium.
\label{statsolfig}}
\end{Figure}

\subsubsection{Inclusion of trap loss}

In our model the bath of thermal atoms is not depleted by the interaction with the 
condensate. Experimentally this can be achieved by replenishing the reservoir by some
mechanism, or by doing the experiment so quickly that the number of particles lost from
the reservoir can be neglected. Furthermore, the calculations presented here for
constant $T$ and $\mu$ remain valid as long as the heat bath parameters $\mu$ and $T$ change 
slowly compared to the time scale of the condensate dynamics.
For the diagonal matrix elements $g_N^0$, we therefore 
obtain a stationary solution different from zero even if we include trap loss. 
Keeping only the leading order terms in $N$ of Eq.~(\ref{matrixeq}), 
we immediately find the Fokker-Planck equation for $g_N^{0}$ to be
\begin{equation}
\dot{g}_N^{0}=\frac{\partial}{\partial N} \{f_1(N) g_N^{0}\} + \frac{1}{2} 
\frac{\partial^2}{\partial N^2} \{f_2(N) g_N^{0}\}
\label{FPEloss}
\end{equation}
where we have defined
\begin{eqnarray}
f_1(N)&=&-2 N W^+(N)+ 2 N (W^-(N)+\gamma_1) \nonumber \\
&& \quad + 4 N^2 \gamma_2(N) + 6 N^3 \gamma_3(N),
\end{eqnarray}
and
\begin{eqnarray}
f_2(N)&=&2 N W^+(N)+ 2 N (W^-(N)+\gamma_1) \nonumber \\
&& \quad + 8 N^2 \gamma_2(N) + 18 N^3 \gamma_3(N).
\end{eqnarray}
The Fokker-Planck equation is valid as long as $\bar{N} \gg 1$ and $\sigma \gg 1$.

We approximate the solution of this Fokker-Planck equation by a Gaussian, and obtain 
the following equation for the mean number of condensate particles $\bar{N}_{\rm loss}$:
\begin{equation}
\bar{f}_1=0.
\label{nloss}
\end{equation}
The width of the Gaussian $\sigma_{\rm loss}$ is approximately given by
\begin{equation}
\sigma_{\rm loss}=\sqrt{\frac{\bar{f}_2}{2 \partial_{\bar{N}_{\rm loss}} 
\bar{f}_1}}.
\end{equation}
Using the assumption $\mu_{\bar{N}_{\rm loss}}/ kT \ll 1$ Eq.~(\ref{nloss}) can be solved 
analytically 
by approximating
\begin{equation}
W^+(N) \approx W^+_a= \frac{4 m (a k T)^2}{ \pi \hbar^3} \mbox{e}^{2 \mu/kT},
\end{equation}
\begin{equation}
W^-(N) \approx W^-_a(N) = W^+_a (1+ \frac{\mu_N}{kT} + \frac{\mu_N^2}{2 k^2 T^2})
\mbox{e}^{-\mu/kT}.
\end{equation}
For the mean number of particles in the condensate, we obtain
\end{multicols}
\Endrule
%\widetext
\begin{equation}
\bar{N}_{\rm loss}= \left(-\frac{\gamma_2(1) + \mu_1 W^+_a \mbox{e}^{-\mu/kT}/2}
{3 \gamma_3(1)+\mu_1^2 W^+_a \mbox{e}^{-\mu/kT}/2} +\sqrt{\left(\frac{\gamma_2(1) + 
\mu_1 W^+_a \mbox{e}^{-\mu/kT}/2}{3 \gamma_3(1)+
\mu_1^2 W^+_a \mbox{e}^{-\mu/kT}/2}\right)^2-\frac{\gamma_1+W^+_a(\mbox{e}^{-\mu/kT}-1)}
{3 \gamma_3(1)+\mu_1^2 W^+_a \mbox{e}^{-\mu/kT}/2}}\right)^{5/2}.
\end{equation}
%\narrowtext
\Beginrule
\begin{multicols}{2}\noindent
Here $\mu_1=\mu_{N=1}/kT$.
Trap loss decreases the number of particles in the condensate. Also, the width of the particle
distribution is decreased by the nonlinear loss.
Both of these effects are well known in nonlinear optics \cite{Helmutnonlin}.
Figure 4 %\ref{traploss}
 shows the effect of trap loss on the mean number of particles in
the condensate for a given $T$ and $\mu$ for the parameters measured in Ref.~\cite{JilaCCC}.
The dominant contribution to the trap loss comes from three-particle loss, while one,- and two-particle losses are almost negligible.
In the following sections we will omit the
subscript loss since all the calculations will be done for finite trap loss.
\begin{Figure}
\infig{5cm}{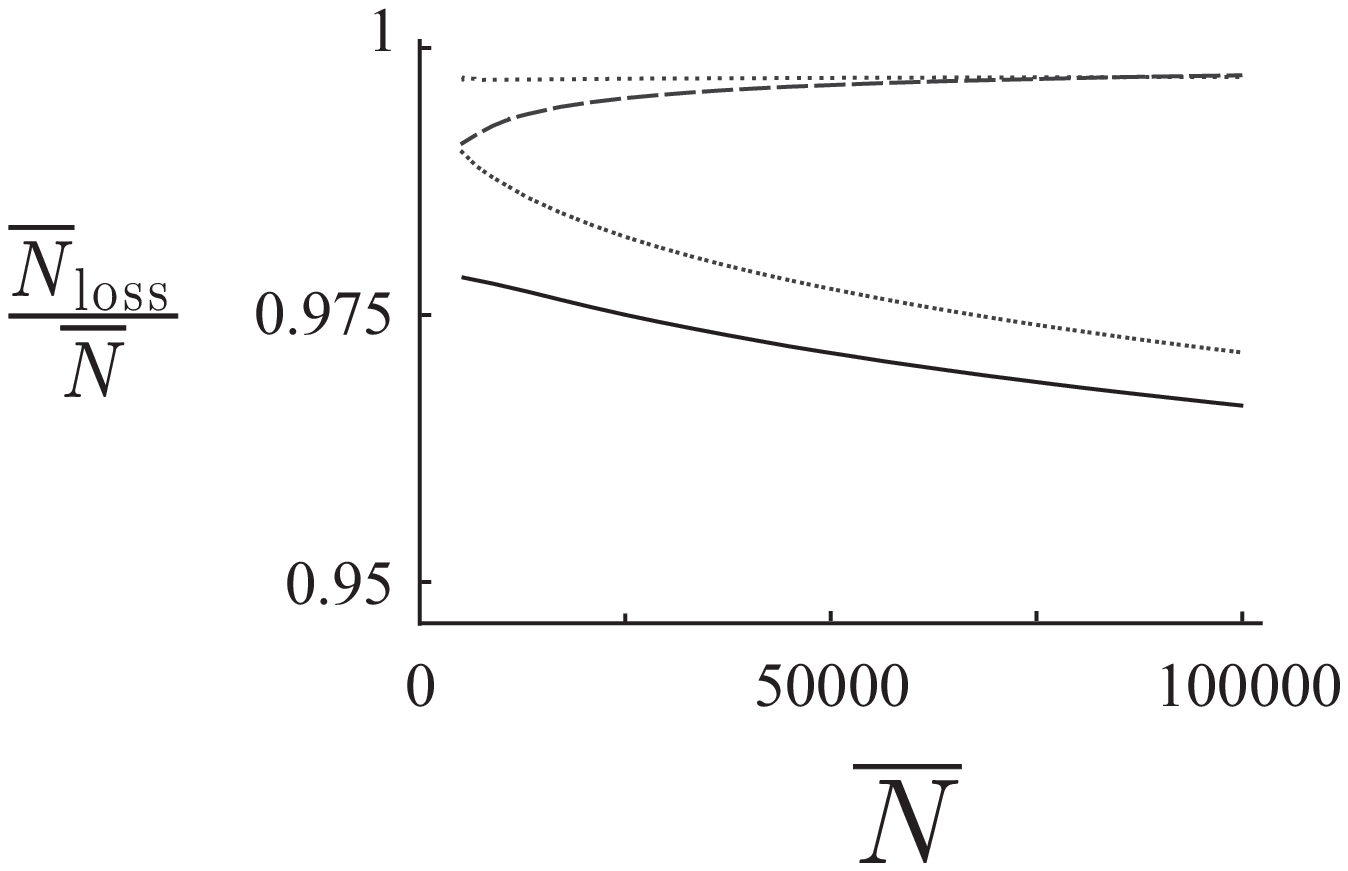}
\Caption{Fig.~4. Influence of trap loss on the mean number of particles in the condensate for 
rubidium.
The trap frequencies are $f_x=f_y=f_z/{\protect \sqrt{8}}=47 {\rm Hz}$. The temperature of the 
thermal
cloud is chosen $T=0.5 \mu K$. One,- two,- and three-particle losses are given by the dotted, 
dashed, and
double dotted lines, respectively. The solid line accounts for all three kinds of loss.
We have used $K_1=1/70 s^{-1}$, $K_2=10^{-22} m^3/s$ and $K_3=6\cdot 10^{-42} m^6/s$ 
{\protect \cite{JilaCCC}}.
\label{traploss}}
\end{Figure}
%Start change
\subsection{Nonstationary solutions}

From Eq.~(\ref{master}) we find the evolution equation \cite{negfluc} for the mean number of 
condensed particles $\bar{N}$.
\begin{eqnarray}
\dot{\bar{N}}&=&2W^+(\bar{N})(\bar{N}+1)-2W^-(\bar{N})\bar{N} \nonumber \\
&& \quad - 2 \gamma_1 \bar{N}-4 \gamma_2(\bar{N}) \bar{N}^2 -6 \gamma_3(\bar{N}) \bar{N}^3.
\label{Nevolution}
\end{eqnarray}
By comparing Eq.~(\ref{Nevolution}) with the results obtained in Ref.~\cite{JilaCCC} we find 
numerical values for the constants $K_i$. 
Equation (\ref{Nevolution}) was investigated in Refs.~\cite{BosGro,roleofquasi} for the case $\gamma_i=0$.
Here we only want to show that trap loss does not change the general behavior of the growth of the
condensate. 
%End change
In Fig.~5 %\ref{growth}
 we show a comparison between the growth curve with and without trap loss. 
For the parameters chosen the time scale of condensate growth is 
influenced only slightly by trap loss.
\begin{Figure}
\infig{5cm}{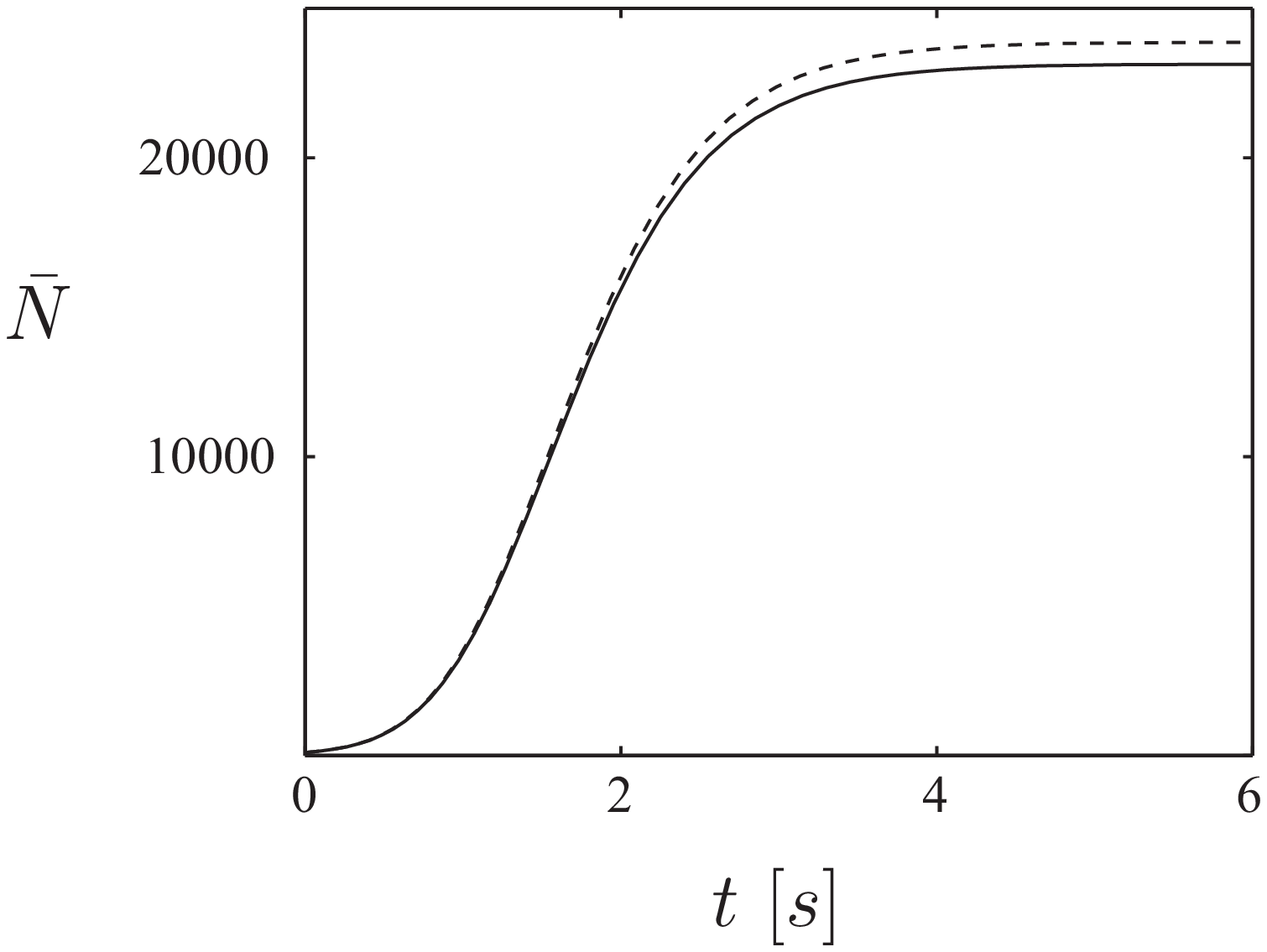}
\Caption{Fig.~5. \protect\label{growth} Comparison between the growth of the condensate with trap loss  
(solid curve) and without trap loss (dashed curve).  The parameters  
are chosen as $K_1=1/70 {\rm s}^{-1}$,
 $K_2=10^{-22} {\rm m}^3/{\rm s}$, and $K_3=6\cdot 10^{-42} {\rm m}^6/{\rm s}$ 
{\protect \cite{JilaCCC}}. Parameters for the thermal particles are 
$T=0.5 \mu {\rm K}$ and $\mu=3 \cdot 10^{-31}{\rm J}$.
The trap frequencies are 
chosen to be $f_x=f_y=f_z/{\protect \sqrt{8}}=47 {\rm Hz}$.  The calculation is done for rubidium. }
\end{Figure}

\subsection{Correlation functions}
\label{corrfct}

In this section we calculate the intensity and amplitude correlation functions for
a condensate in the stationary state found in 
Sec.~\ref{statsolsec}.

\subsubsection{Ito equation}

First we will show that if we restrict ourselves to situations where a large number
of particles occupies the condensate and the density operator is almost diagonal in the
basis $|N \rangle_N$, we can approximate
Eq.~(\ref{master}) by a master equation of standard Lindblad form.
To do so, we consider terms of the form
\begin{equation}
2 C^{\dagger} \{F(\hat{N}) \rho \} C - [C C^{\dagger},\{F(\hat{N}) \rho \}]_+,
\label{comp1}
\end{equation}
define the operator $D=\sqrt{F(\tilde{N})} C$, and compare Eq.~(\ref{comp1}) with
\begin{equation}
2 D^{\dagger} \rho D - [D D^{\dagger}, \rho]_+
\label{comp2}
\end{equation}
for a matrix element $g_N^{\nu}$.
Since we assume the trap loss to be a small effect compared to the interaction of
the condensate with the thermal particles, the terms $C^{\dagger}C F(\hat{N})$ of the
master equation (\ref{master}) are of order $W^+\bar{N}$. The only exception are the
terms involving $R_{00}$ \cite{R00just}.
Therefore, we find for all the terms in our master equation that the difference of 
Eqs.~(\ref{comp1}) and (\ref{comp2}) for a matrix element $g_N^{\nu}$ is of the order of
$\bar{N} W^+(\bar{N}) (\nu/\bar{N})^2$. We will linearize the equations and neglect
all terms of order $W^+(\bar{N})/\bar{N}$. Therefore, we approximate the master equation
(\ref{master}) by replacing all the terms of the form Eq.~(\ref{comp1}) by expressions of 
the 
form of Eq.~(\ref{comp2}). This enables us to write down the Ito equation straightforwardly 
for the evolution of the system operators in the Heisenberg picture.
Note that the noise
terms appearing in the Ito equations do not have 
a direct physical interpretation. 
We only need
them to have a mathematical equivalence between the solutions of the master equation and
the solutions of the Ito equation. 
The Ito stochastic equation for an operator $X$ in the Heisenberg picture reads
\end{multicols}
\Endrule
%\widetext
\begin{eqnarray}
d X & = & \Bigl\{ \frac{i}{\hbar} \left[ H_0,X \right]  
   -[X,\sqrt{W^+}B B^{\dagger}\sqrt{W^+}]_+ + 2 \sqrt{W^+}B X B^{\dagger}\sqrt{W^+}
   -[X,\sqrt{W^-+\gamma_1} B^{\dagger} B \sqrt{W^-+\gamma_1}]_+ + \nonumber \\
&& 2 \sqrt{W^-+\gamma_1}
   B^{\dagger} X B \sqrt{W^-+\gamma_1} 
   -[X,\sqrt{R_{00}} (B B^{\dagger})^2  \sqrt{R_{00}}]_+ + 2 \sqrt{R_{00}}B B^{\dagger} X B 
   B^{\dagger} \sqrt{R_{00}} \nonumber \\
&& -[X,\sqrt{\gamma_2} B^{\dagger} B^{\dagger} B B \sqrt{\gamma_2}]_+ + 2 \sqrt{\gamma_2}
   B^{\dagger} B^{\dagger} X B B \sqrt{\gamma_2}
   -[X,\sqrt{\gamma_3} B^{\dagger} B^{\dagger} B^{\dagger} B B B \sqrt{\gamma_3}]_+ + 
   2 \sqrt{\gamma_3} B^{\dagger} B^{\dagger} B^{\dagger} X B B B \sqrt{\gamma_3} \Bigr\} dt  
\nonumber \\
&& -\left([X,\sqrt{2 W^+}B] dC_1-[X,B^{\dagger} \sqrt{2 W^+}] dC_1^{\dagger}\right)
   -\left([X,\sqrt{2 (W^-+\gamma_1)}B^{\dagger}] dC_2-[X,B \sqrt{2 (W^-+\gamma_1)}] 
   dC_2^{\dagger}\right) \nonumber \\
&& -\left([X,\sqrt{2 R_{00}}B B^{\dagger}] dC_3-[X,\sqrt{2 R_{00}} B B^{\dagger}]
   dC_3^{\dagger}\right)
   -\left([X,\sqrt{2 \gamma_2} B^{\dagger} B^{\dagger}] dC_4-[X,B B \sqrt{2 \gamma_2}] 
   dC_4^{\dagger}\right) \nonumber \\
&& -\left([X,\sqrt{2 \gamma_3} B^{\dagger} B^{\dagger} B^{\dagger}] dC_5-[X,B B B 
\sqrt{2 \gamma_3}] dC_5^{\dagger}\right), \nonumber \\
\label{XIto}
\end{eqnarray}
%\narrowtext
\Beginrule
\begin{multicols}{2}\noindent
where $dC_i$ are Ito noise increments.
The only expectation values that are different from zero are
\begin{equation}
\langle dC_i(t) dC_i^{\dagger}(t) \rangle=dt.
\end{equation}
Note that all the rates appearing in Eq.~(\ref{XIto}) depend on the number operator 
$\tilde{N}$ and, that
therefore we use relations like 
[for example, for $W^-(\tilde{N})$]
\begin{eqnarray}
B W^-(\tilde{N})&=&W^-(\tilde{N}+1) B \approx \left(W^-(\tilde{N})+ 
\frac{d W^-(\tilde{N})}{d \tilde{N}} \right) B \nonumber \\
&=& (W^-+\partial_{\tilde{N}} W^-)B
\end{eqnarray}
to calculate commutators between these rates and $B$.

\subsubsection{Intensity fluctuations}
\label{meannumber}
We define the operator
\begin{equation}
\delta I_B=\frac{\tilde{N}-\langle \tilde{N} \rangle}{\sqrt{\langle \tilde{N} \rangle}},
\end{equation}
where we omit the time dependence whenever this can be done without causing confusion.
For $\delta I_B$ we obtain
\begin{equation}
d \delta I_B = - \gamma_I \delta I_B dt+dC_I,
\label{intfluc}
\end{equation}
where
\begin{eqnarray}
dC_I &=& \left(\sqrt{2W^+}B dC_1+B^{\dagger} \sqrt{2W^+} dC_1^{\dagger}+ \right. \nonumber \\
& & -\sqrt{2(W^-+\gamma_1)}B^{\dagger} dC_2-B \sqrt{2(W^-+\gamma_1)} dC_2^{\dagger} 
\nonumber\\
& &  - 2\sqrt{2\gamma_2}B^{\dagger}B^{\dagger} dC_4-2 B B \sqrt{2\gamma_2} dC_4^{\dagger} 
\nonumber \\
& &  \left. -3\sqrt{2\gamma_3}B^{\dagger}B^{\dagger}B^{\dagger} dC_5-3 B B B \sqrt{2\gamma_3} 
dC_5^{\dagger} \right) \frac{1}{\sqrt{\bar{N}}}.\nonumber\\
\end{eqnarray}
We expand around $\bar{N}$ and obtain
\begin{equation}
\gamma_I = \partial_{\bar{N}} \bar{f}_1 
\approx  2 \bar{N} \partial_{\bar{N}} \bar{W}^{-}+\frac{8}{5} \bar{\gamma}_2\bar{N}
 +\frac{24}{5} \bar{\gamma}_3 \bar{N}^2.
\end{equation}
which is now only a function of the expectation value of $\tilde{N}$.
Keeping only these highest-order terms, we can solve  Eq.~(\ref{intfluc}) and obtain
\begin{equation}
\delta I_B(t)=\delta I_B(0) \mbox{e}^{-\gamma_I t} + \int_0^t \mbox{e}^{-\gamma_I (t-t')} \, 
dC_I(t').
\end{equation}
Using $\langle dC_I(t) dC_I^{\dagger}(t)\rangle = \bar{f}_2 dt /\bar{N}$
and considering only $t+s \gg \gamma_I^{-1}$ we obtain                 
for the  correlation function $\langle \tilde{N}(t), \tilde{N}(s) \rangle$  

\begin{equation}
\langle N(t), N(s) \rangle=\frac{\bar{f}_2}{2 \gamma_I} 
\mbox{e}^{-\gamma_I \left|t-s\right|}=\sigma^2\mbox{e}^{-\gamma_I \left|t-s\right|}.
\end{equation}
Note that the operator $\delta I_B$ is of order $1$.
For $\gamma_2=\gamma_3=0$ we find for the Mandel $Q$ parameter
\begin{equation}
Q \approx \frac{ 5 kT}{2 \mu_{\bar{N}}}-1.
\end{equation}
This means that as long as there is a significant thermal bath, $Q$ will always be larger than
$0$ since in this case $\mu_{\bar{N}}/kT < 1$ holds. However, since the result for $Q$ also
follows from statistical mechanics the restriction $\mu_{\bar{N}}/kT < 1$ is not necessary for 
this 
result to be valid, and we obtain sub-Poissonian statistics for $\mu_{\bar{N}}/kT > 5/2$.

\subsubsection{Amplitude fluctuations}
We use the Ito equation introduced above to calculate the phase fluctuations. In 
particular we calculate the correlation function
\begin{equation}
\langle B^{\dagger}(t) B(s) \rangle.
\end{equation}
First we simplify Eq.~(\ref{XIto}) for $X=B$ and find
\begin{equation}
dB= -\gamma_B B dt + dC_B,
\label{BIto}
\end{equation}
where to leading order in $\bar{N}$
\begin{equation}
\gamma_B = \frac{i \mu_{\bar{N}}}{\hbar}+ \frac{16}{25} \bar{R}_{00}+ 
\sqrt{\bar{N}} \delta I_B \left( \frac{i \partial_{\bar{N}} \mu_{\bar{N}}}{\hbar} 
+ \frac{\gamma_I}{2 \bar{N}} - \frac{32 \bar{R}_{00}}{125 {\bar{N}}}\right).
\label{gab}
\end{equation}
The terms of order $\bar{W}^{\pm}/{\bar{N}}$ have been neglected. 
Expressions like Eq.~(\ref{gab}) appear in optics in connection with the
Kerr effect \cite{Kerr}. The noise term reads
\end{multicols}
\Endrule
%\widetext
\begin{eqnarray}
dC_B &=&  \left(\sqrt{2 W^+} + \frac{\partial_N W^{+}}{\sqrt{2 W^+}}
   B^{\dagger}B\right) dC_1^{\dagger} -\frac{\partial_N W^{+}}{\sqrt{2 W^+}} BB dC_1 +
   \left(-\sqrt{2 (W^-+\gamma_1)} - \frac{\partial_N W^{-}}{\sqrt{2 (W^-+\gamma_1)}}
   B B^{\dagger}\right) dC_2  \nonumber \\
&& +\frac{\partial_N W^{-}}{\sqrt{2 (W^-+\gamma_1)}} BB
   dC_2^{\dagger}
   + \left( - \frac{\partial_N R_{00}}{\sqrt{2 R_{00}}} B^{\dagger} B - \sqrt{2 R_{00}} 
\right) B
   \left( dC_3-dC_3^{\dagger} \right)
   + \left( - \frac{\partial_N \gamma_2}{\sqrt{2 \gamma_2}} B^{\dagger} B^{\dagger} B - 
   2 \sqrt{2 \gamma_2} B^{\dagger} \right) dC_4  \nonumber \\
&& +\frac{\partial_N \gamma_2}{\sqrt{2 \gamma_2}} BBB dC_4^{\dagger}
   + \left( - \frac{\partial_N \gamma_3}{\sqrt{2 \gamma_3}} B^{\dagger} B^{\dagger} 
B^{\dagger} B - 
   3 \sqrt{2 \gamma_3} B^{\dagger} B^{\dagger} \right) dC_5 + \frac{\partial_N \gamma_3}{\sqrt{2 
\gamma_3}} 
   BBBB dC_5^{\dagger}.
\end{eqnarray}
%\narrowtext
\Beginrule
\begin{multicols}{2}\noindent
We define the operator
$\Phi=\frac{1}{\sqrt{B^{\dagger}B+1}}B$
and find the following equation for $\Phi$
\begin{equation}
d\Phi= \left(-\gamma_B dt + dC_{\phi}- \frac{1}{2 \sqrt{\bar{N}}} d \delta I_B\right) \Phi
\label{dPhi}
\end{equation}
Keeping only the leading terms, we obtain for $dC_{\phi}$
\begin{equation}
dC_{\phi} = \left( - \frac{\partial_N R_{00}}{\sqrt{2 R_{00}}} B^{\dagger} B - \sqrt{2 R_{00}} 
\right)
\left( dC_3-dC_3^{\dagger} \right)
\end{equation}
and find
\begin{equation}
dC_{\phi}(t) dC_{\phi}(t)= 2\left(- \frac{16}{25} \bar{R}_{00} + 
\sqrt{\bar{N}} \delta I_B \frac{32 \bar{R}_{00}}{125 {\bar{N}}}\right) dt.
\end{equation}
Equation (\ref{dPhi}) can be treated as a $c$-number equation for $\Phi$, since $i$ times the
noise term in this equation has the properties of a classical noise term. We define an
operator $\Phi=\mbox{e}^{i \phi}$ and obtain, for its increment,
\begin{equation}
d\phi=\left(-\frac{\mu_{\bar{N}}}{\hbar} - \frac{\partial_{\bar{N}} \mu_{\bar{N}}}{\hbar} 
\sqrt{{\bar{N}}} \delta I_B
\right) dt -idC_{\phi}
\label{dphi}
\end{equation}
In this equation the intensity noise $dC_I$ is independent of the phase noise $dC_{\phi}$,
so that the 
equation for $\phi$ is very similar to the equations appearing in the 
phase diffusion model for a laser with colored noise \cite{colornoise}.
We may use the same method as used in quantum optics to calculate the correlation function 
$\langle B^{\dagger}(t) B(s) \rangle$, and obtain
\begin{eqnarray}
&&\langle B^{\dagger}(t) B(s) \rangle = {\bar{N}} \mbox{e}^{\left(  
i\frac{\mu_{\bar{N}}}{\hbar}
 - \frac{16}{25} R_{00} \right) \left( t-s \right)} \nonumber \\
&& \times \mbox{e}^{\left(-\left(\frac{\sigma \partial_{\bar{N}} \mu_{\bar{N}}}{\gamma_I 
\hbar}\right)^2 
\left( \gamma_I (t-s)+\mbox{e}^{-\gamma_I (t-s)} -1\right)\right)}
\label{BdBeq}
\end{eqnarray}
for $t>s$ in the stationary state. In the range of validity of our approximations
the result for this correlation function Eq.~(\ref{BdBeq}) agrees with the result obtained 
in QKIII [Eq.~(184) of Ref.~\cite{QKIII}].
If we were to keep the terms of $O(1/\sqrt{{\bar{N}}})$ in the
noise terms the phase noise would
no longer be independent of the intensity noise and the 
solution to Eq.~(\ref{dphi}) would not be so easy to find. 
The correlation function (\ref{BdBeq}) depends only on the time difference $t-s$. 
The spectrum is found by a Fourier transformation in the time difference $t-s$
\begin{equation}
S(\omega)=\frac{1}{\sqrt{2 \pi}} \int_{-\infty}^{\infty} d\Delta t \,  \mbox{e}^{-i \omega 
\Delta t}
\langle B^{\dagger} (t+\Delta t) B(t) \rangle.
\label{Speceq}
\end{equation}
\begin{Figure}
\infig{5cm}{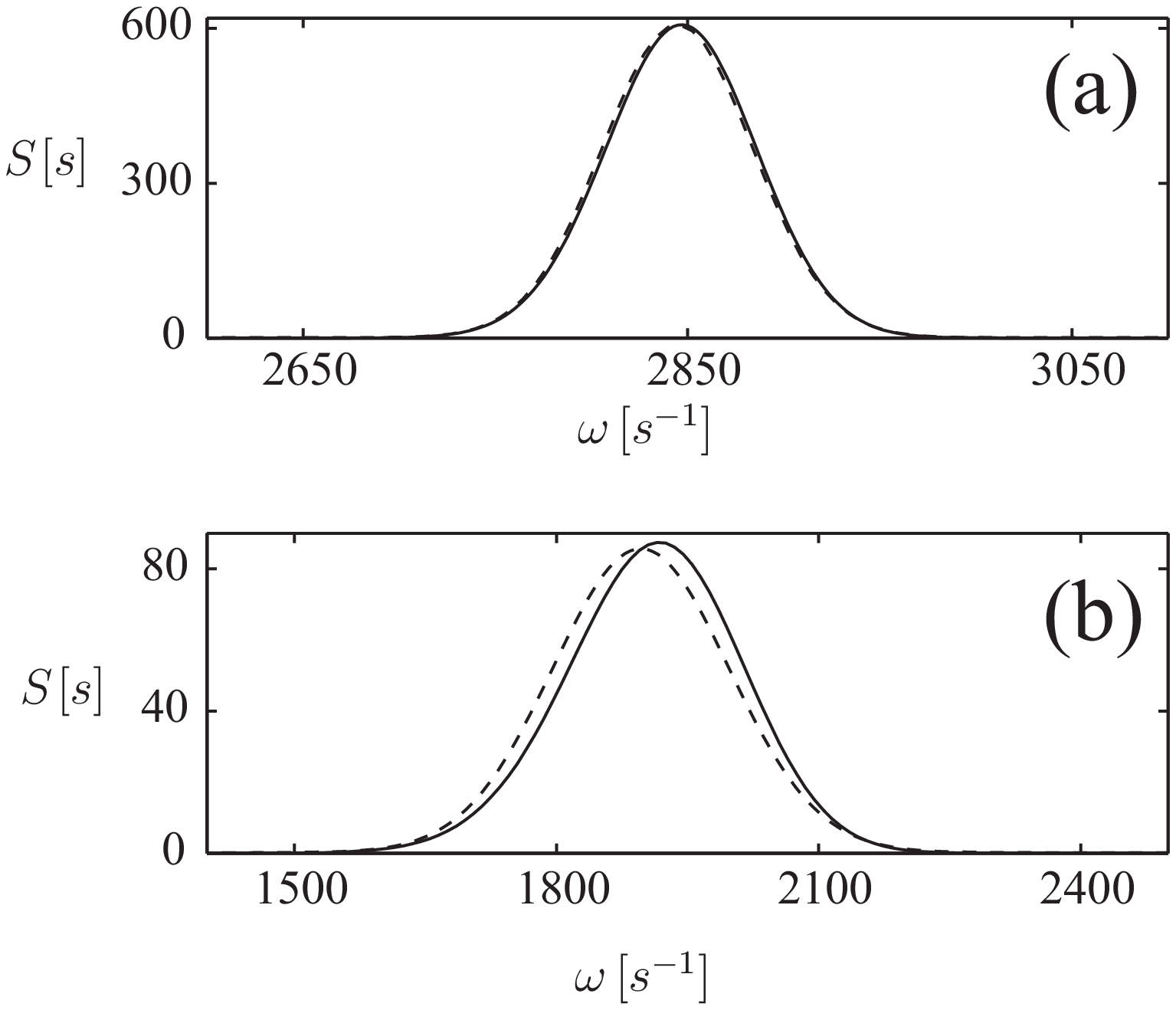}
\Caption{Fig.~6. Spectrum of the amplitude fluctuations $S(\omega)$ as defined in Eq.~(\ref{Speceq}) 
against $\omega$ for rubidium. The trap frequencies 
are chosen to be $f_x=f_y=f_z/{\protect \sqrt{8}}=47 {\rm Hz}$.
The numerical solutions are given by the solid lines, and the dashed lines represent the
analytical results.
In (a) the parameters of the thermal cloud are chosen to be $T=0.25 \mu K$ and $\mu=3 \cdot
10^{-31} J$. The mean number of particles therefore is $\bar{N}=23800$. Plot (b) shows the
spectrum for $T=0.9 \mu K$ and $\mu=2 \cdot 10^{-31} J$, and therefore a mean number of 
condensate
particles $\bar{N}=8640$. \label{specfig}}
\end{Figure} 
Fig.~6 %\ref{specfig}
 shows a comparison between the analytic result [Eq.~(\ref{BdBeq})]
and the direct numerical solution of Eq.~(\ref{master}). For a large number of condensate 
particles
$\bar{N}$, the results agree very well with each other. In case of small particle numbers 
$\bar{N}$ the 
linearization used to obtain the analytic result shifts the curve compared to the numerical 
result. Even so,
the shape of the solution is very well approximated by the analytic formula.
The spectrum $S(\omega)$ is expected to be of Lorentzian shape
around its maximum value. Further away from the center, the shape becomes Gaussian.
However, for the parameters chosen in Fig.~6 %\ref{specfig}
 the Gaussian part dominates.

%
% ********************************************************************************
% ***                                                                          ***
% ***                           Conclusions                                    ***
% ***                                                                          ***
% ********************************************************************************

\section{Conclusions}

In this paper we have calculated the correlation functions for amplitude
(phase) and intensity fluctuations of a Bose condensate due to
interactions with a heat and particle reservoir, representing
uncondensed atoms at finite temperature. The present analysis is valid
for a strongly condensed system confined in a trapping potential,
ignoring contributions from quasiparticle excitations \cite{QKIII}. 
Finally, we point out that the present theory is readily adapted to a
class of highly interesting problems, such as the study of decoherence
in Josephson-like situations, where two trapped condensates are brought
into contact and the quantum dynamics of the relative phase of the two
condensates is observed \cite{Villain,Sols}. 

\acknowledgements
We like to thank M.~Holland, J.~Williams, K.~Ellinger and H.~Ritsch for stimulating 
discussions. 
This work was supported by the Marsden Fund under Contract No.~PVT-603, and
by {\"O}sterreichische Fonds zur F{\"o}rderung der wissenschaftlichen Forschung.
Part of this work was supported by TMR Network No. ERB 4061 PL 95-0044.

%
% ********************************************************************************
% ***                                                                          ***
% ***                            Appendix                                      ***
% ***                                                                          ***
% ********************************************************************************

\appendix
%Start change
\section{Stationary solution for a canonical bath of particles}
\label{cancorr}

We want to investigate the fluctuations in the number of particles of a Bose-Einstein 
condensate, assuming the system to be in the canonical ensemble. We will include 
interactions between the Bose particles in our analysis, and investigate their effects 
on the condensate fluctuations. On first glance one might expect that it would 
be satisfactory to account for the interaction effects by just including the mean field 
of the condensate (in the Thomas-Fermi approximation). However, in this approach fluctuations
in the size of the condensate lead to an unrealistically large shift of the energy levels,
which prevents any condensation \cite{QKV}. The inclusion of the mean field of the thermal
density of particles reduces the shift of the energy levels. As long as fluctuations 
in the chemical potentials of the thermal cloud $\mu$ and the condensate $\mu_N$ are small 
compared to the energy gap between the condensate energy and the first  
excitation energy, fluctuations will not lead to degeneracy in the thermal cloud. The 
excitation energies therefore depend mainly on the total number of 
thermal particles $M$ and the number of particles in the condensate $N$.
We will assume that the eigenenergies of the excited levels depend on $M$ and $N$ but 
are independent of $n_m$, i.e., the microstate of the system. As shown in 
Ref.~\cite{Stringari2}, the density of particlelike states is much larger than the density of 
quasiparticles. Thermodynamic quantities of the Bose gas are therefore mainly determined 
by the particlelike states, which allows us to use the Hartree-Fock approximation for 
describing the interactions in the thermal cloud.

\subsection{Excited modes}
\label{exmodes}

We denote the energies and the wave functions of the excited states by $\epsilon_m$ and
$\xi_m({\bf x})$, respectively. The occupation of these levels is $n_m$.
In the Hartree-Fock approximation \cite{HFapprox}, the effective potential 
for the thermal particles $V_{\rm eff}({\bf x})$ is given by
\begin{equation}
V_{\rm eff}({\bf x})= V_T({\bf x}) + 2 u N \left | \xi_N({\bf x})\right|^2+2 u 
\tilde{n}({\bf x}),
\label{Veff}
\end{equation}
with $\tilde{n}({\bf x})$ the density of the noncondensed particles.

\subsection{Weakly interacting Bose gas in the canonical ensemble}

The density operator of a Bose gas in the canonical ensemble is given by 
\begin{equation}
\rho_c=\frac{1}{Z_c} \mbox{e}^{-\beta H},
\label{densop}
\end{equation}
where $\beta=1/kT$, with $T$ the temperature of the system. The partition 
function $Z_c$ is given by $Z_c=\mbox{tr}\left\{\mbox{e}^{-\beta H} \right\}$.
We want to investigate the properties of a Bose condensate in the canonical
ensemble. The eigenstates $\{\xi_N({\bf x}),\xi_m({\bf x})\}$ and the corresponding 
eigenenergies $\{E_0,\epsilon_m\}$ depend on the number of condensate particles $N$ 
and on the number of particles out of the condensate $M=\sum_m n_m$. The total number 
of particles $N_{\rm tot}$ is constant,
\begin{equation}
N_{\rm tot}=N+M=\rm{const}.
\end{equation}
The state of the system with $N$ particles in the condensate and $n_m$ particles
in the levels $\epsilon_m$ is denoted by $\left | {\bf n} \right. \rangle$ where
${\bf n}=\{N,n_1,...,n_m,...\}$.
We can therefore write for the matrix elements of the density operator 
$p({\bf n})=\langle {\bf n} \left| \rho_c \right  | {\bf n} \rangle$
\begin{equation}
p({\bf n}) \propto  \mbox{e}^{-\beta E_0(N,M) -\beta \sum_m \epsilon_m(N,M) n_m}
\label{eqpn}
\end{equation}
As can be seen from Eq.~(\ref{eqpn}), the condensate energy $E_0(N,M)$ and the 
excitation energies $\epsilon_m(N,M)$ are functions of the number of condensate particles
$N$ and the number of noncondensed particles $M$. In our calculations we will assume
that the influence of the number of noncondensed particles $M$ on the condensate energy
is negligible, since the number of noncondensed particles that are inside the condensate region
is much smaller than the number of condensed particles. Moreover, the width of the condensate
particle distribution is only influenced by the change of the number of noncondensed particles 
in the condensate region due to fluctuations in $M$. The other interaction effects
i.e., (i) the influence of the condensate mean field on the excited levels, (ii) the influence 
of the mean field of the thermal cloud on the excited levels, and (iii) the influence of the
condensate mean field on the energy of the condensate, will be included in our calculations.

\subsection{Particle number distribution of the condensate}

Since we are only interested in the probability of finding $N$ particles in the condensate
we want to find
\begin{equation}
p(N,M)=\sum_{\{n_m\}} p(N,n_m)
\end{equation}
summed under the constrained $M=\sum_m n_m$.
We can do this summation by introducing a contour integral writing 
\begin{eqnarray}
p(N,M) &\propto& \mbox{e}^{-\beta E_0(N,M)} \\
&\times&\frac{1}{2 \pi i} \int_C \frac{dz}{z} z^{-M} 
\prod_m \sum_{n_m=0}^{\infty} \mbox{e}^{-\beta \epsilon_m(N,M) n_m} z^{n_m}, \nonumber 
\end{eqnarray}
and integrating using the method of steepest descents to obtain
\begin{eqnarray}
p(N,M) &&\propto \mbox{e}^{-\beta E_0(N,M)-(M+1) \ln(\bar{z})-\sum_m \ln\left(1-\bar{z}
\mbox{e}^{-\beta \epsilon_m(N,M)}\right)} \nonumber \\
\times&& \left(\frac{M+1}{\bar{z}^2}+\sum_m \frac{\mbox{e}^{-2\beta \epsilon_m(N,M)}}
{\left(1-\bar{z}\mbox{e}^{-\beta \epsilon_m(N,M)}\right)^2}\right)^{-1/2}.
\label{Partdist}
\end{eqnarray}
$\bar{z}$ depends on $M$ and $N$, and is given by the solution of 
\begin{equation}
M+1=\sum_m \frac{1}{\mbox{e}^{\beta \epsilon_m(N,M)} \bar{z}^{-1}-1}.
\label{Mdef}
\end{equation}
By defining
\begin{eqnarray}
F(N,M)&=&E_0(N,M)+\frac{M}{\beta} \ln(\bar{z})\nonumber \\
&& + \frac{1}{\beta} \sum_m \ln\left(1-\bar{z} \mbox{e}^{-\beta \epsilon_m(N,M)}\right),\label{Adef} \\
C(N,M)&=&\left\{ \ln\left(\partial_M \ln(\bar{z})\right)
- \ln\left(\bar{z} \sum_m \frac{\beta \partial_M \epsilon_m(N,M)}{\mbox{e}^
{\beta \epsilon_m(N,M)}-\bar{z}} \right. \right. \nonumber \\
&& + \left. \left. \bar{z}^2 \sum_m \frac{\beta \partial_M \epsilon_m(N,M)}
{\left(\mbox{e}^{\beta \epsilon_m(N,M)}-\bar{z}\right)^2}+1\right) \right\}\frac{1}{2},
\end{eqnarray}
and using Eq.~(\ref{Mdef}), we may rewrite Eq.~(\ref{Partdist})
\begin{equation}
p(N,M) \propto  \mbox{e}^{-\beta F(N,M)+C(N,M)}.
\label{Partdistn}
\end{equation}
$C(N,M)$ is a small logarithmic correction to $\beta F(N,M)$ that we neglect in
our further calculations.
Consistent with the neglect of $C(N,M)$, we approximate $M+1$ by $M$ in 
Eq.~(\ref{Mdef}) and use the equations
\begin{equation}
M=\sum_m \frac{1}{\mbox{e}^{\beta \epsilon_m(N,M)} \bar{z}^{-1}-1},
\label{Mdefapp}
\end{equation}
and
\begin{equation}
p(N,M) \propto  \mbox{e}^{-\beta F(N,M)}.
\label{Partdistnapp}
\end{equation}
$F(N,M)$ is the Helmholtz free energy of the system. The chemical potentials
of the condensate and the thermal bath are therefore given by the partial 
derivatives of $F(N,M)$ with respect to $N$ and $M$, respectively.

\subsubsection{Stationary solution}
The chemical potential of the condensate in the canonical ensemble $\mu_c(N,M)$ is
given by
\begin{eqnarray}
\mu_c(N,M)&=&\partial_N F(N,M) \nonumber \\
&=&\partial_N E_0(N,M)+\sum_m \frac{\bar{z} \partial_N \epsilon_m(N,M)}{\mbox{e}^{\beta 
\epsilon_m}-\bar{z}},
\label{muccan}
\end{eqnarray}
and the chemical potential of the noncondensed particles in the canonical ensemble
$\mu_v(N,M)$ reads
\begin{eqnarray}
\mu_v(N,M)&=&\partial_M F(N,M) \nonumber \\
&=&\frac{\ln(\bar{z})}{\beta}+\partial_M E_0(N,M) +\nonumber \\
&&\sum_m \frac{\bar{z} \partial_M \epsilon_m(N,M) }
{\mbox{e}^{\beta \epsilon_m}-\bar{z}} 
\label{muvcan}
\end{eqnarray}
If the total number of particles is fixed, the condensate particle distribution will reach its 
maximum value for $N=\bar{N}$ and $M=\bar{M}=N_{\rm tot}-\bar{N}$, where $\bar{N}$
can be found by solving
\begin{equation}
\mu_v(\bar{N},N_{\rm tot}-\bar{N})=\mu_c(\bar{N},N_{\rm tot}-\bar{N}). 
\label{mueq}
\end{equation}
The width of the stationary canonical distribution $\sigma_{\rm can}$ is given by
\begin{equation}
\sigma_{\rm can}=\left\{\beta \partial_{\bar{N}} \left( \mu_c(N,N_{\rm tot}-N) - 
\mu_v(N,N_{\rm tot}-N)\right)\right\}^{-1/2}.
\end{equation}
In the above equations (\ref{muccan}) and (\ref{muvcan}), the terms
including derivatives of $\epsilon_m(N,M)$ are corrections to the chemical potentials due to
the shift of the excited energy levels with changing $M$ and $N$. 
The term $\partial_M E_0(N,M)$ in Eq.~(\ref{muvcan}) is the correction to the chemical potential of
the thermal bath due to changes in the condensate energy. By assuming $E_0(N,M)\approx E_0(N)$, we
will neglect this term.

\subsubsection{Method of solving for $p(N,M)$}
For simplicity we will assume that the energy and wave function of the condensate are
only functions of the number of condensed particles, and use the Thomas-Fermi
expressions (\ref{condxi}) and (\ref{condE0}), respectively. The influence of the
thermal particles on the energy and wave function of the condensate will become
important at temperature close to $T_c$ and may therefore only be neglected at very
low temperatures \cite{QKV}.

We replace the sum in Eq.~(\ref{Mdefapp}) by an integral 
\begin{equation}
\sum_m  \rightarrow \int d \epsilon \, g(\epsilon),
\end{equation}
where $g(\epsilon)$ is the density of states and use the semiclassical expression
\begin{equation}
g(\epsilon)=\frac{1}{(2 \pi \hbar)^3} \int d^3x \, \int d^3p \, \delta\left(\epsilon-\frac{{\bf p}^2}{2m}-
V_{\rm eff}({\bf x})\right),
\end{equation}
to obtain, for a spherically symmetric effective potential and given $M$ and $N$,
\begin{eqnarray}
\label{Msol}
M&=&4\pi \int_0^{\infty} r^2 dr \, \tilde{n}(r) = \\
 &=&\left(\frac{m k T}{2 \pi \hbar^2}\right)^{3/2} 4 \pi \int_0^{\infty} r^2 dr \, G_{3/2}
\left(\frac{V_{\rm eff}({\bf x})}{kT}-\ln(\bar{z})\right). \nonumber
\end{eqnarray}
Here $G_{\sigma}(x)=\sum_{n=1}^{\infty} n^{-\sigma} \mbox{e}^{-n x}$
denotes the Bose functions. By solving this equation numerically we obtain $\bar{z}$ as
well as the density of noncondensed particles $\tilde{n}(r)$. 
Given a fixed total number of particles $N_{\rm tot}$ and a temperature $T$ of the system,
we find the particle distribution function $p(N,M)$ from Eq.~(\ref{Partdistnapp}) by replacing
\begin{eqnarray}
-\sum_m \ln\left(1-\bar{z}\mbox{e}^{-\beta \epsilon_m(N,M)}\right) && \\
=\left(\frac{2 m k T}{\pi \hbar^2}\right)^{3/2} \frac{\pi}{2}
\int_0^{\infty} r^2 dr &&\, G_{5/2}\left(\frac{V_{\rm eff}({\bf x})}{kT}-\ln(\bar{z})\right). \nonumber
\end{eqnarray}
in the expression Eq.~(\ref{Adef}) for $F(N,M)$.

\subsubsection{Solutions}

We want to compare $p(N,M)$ with the solution for a grand canonical bath of thermal particles not 
influenced by mean-field effects as given in Eq.~(\ref{Gaussappr}). 
To distinguish between the effect of choosing a different thermodynamic ensemble and the effect of
including the change of the mean field due to condensate fluctuations, we will plot a third distribution
obtained from the canonical ensemble with a fixed mean effective potential $\bar{V}_{\rm eff}({\bf x})$
that is given by
\begin{equation}
\bar{V}_{\rm eff}({\bf x})= V_T({\bf x}) + 2 u \bar{N} \left | \xi_{\bar{N}}({\bf x})\right|^2+
2 u \bar{n}({\bf x}),
\label{Veffsim}
\end{equation}
where $\bar{n}({\bf x})$ is the thermal density obtained for $\bar{N}$ particles in the condensate, 
and $N_{\rm tot}-\bar{N}$ particles in the thermal cloud. 
\begin{Figure}
\infig{5cm}{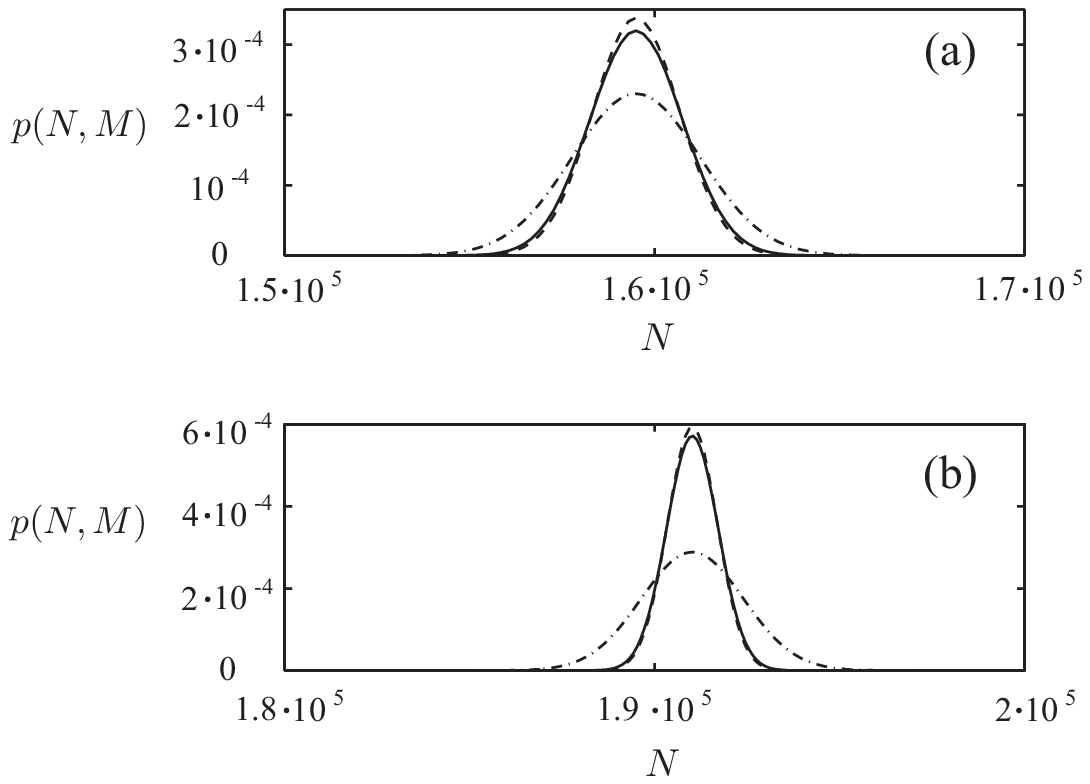}
\Caption{Fig.~7. Condensate particle distribution for (a) $T=0.35 \mu K$, $\bar{N}=1.595 \cdot
10^5$, and $\bar{N}/N_{\rm tot} \approx 10 \%$, and (b) $T=0.2 \mu K$, $\bar{N}=1.910 \cdot 10^5$, and 
$\bar{N}/N_{\rm tot} \approx 33 \%$. Both plots are for Rb  and a trap
frequency of $f_x=f_y=f_z=66 \rm{Hz}$. The solid line shows the solution for the canonical ensemble 
including the mean-field effects, the dashed line is the canonical solution using the mean effective
potential $\bar{V}_{\rm eff}({\bf x})$, and the dash-dotted line is the grand canonical result. 
\label{probdis}}
\end{Figure} 
In Fig.~7 %\ref{probdis}
, we plot the different particle distributions.
For the three curves in each set the number of particles in the condensate $\bar{N}$ is 
the same. Since the inclusion of the mean-field effects shifts the chemical potential of
the noncondensed atoms, the two canonical 
solutions are not plotted for the same total number of particles. If we were to plot
the canonical solutions with and without inclusion of the mean-field effects at the same
total number of particles $N_{\rm tot}$, their maxima would be shifted against each other. 
As can be seen from Fig.~7 %\ref{probdis}
, the stationary state of a Bose condensate in the canonical ensemble is different
from the stationary state of a Bose condensate interacting with a bath of particles in 
the grand canonical ensemble.
When the ratio $\bar{N}/N_{\rm tot} \approx 10\%$ this difference is not significant, but
for larger condensates, i.e., $\bar{N}/N_{\rm tot} > 30\%$, the difference might become substantial.

As shown in this appendix, the model used to describe the thermal bath of atoms may have some
influence on the particle fluctuations in the condensate. We also expect the intensity relaxation
rate $\gamma_I$ to be influenced by the particular model used for the noncondensed particles. However,
a further more detailed analysis of such effects including a treatment of the excitation modes
as given in Ref.~\cite{Griffin} lies beyond the scope of this paper, and will be
presented elsewhere \cite{QKV,QKVI}.

\section{Calculation of $R_{00}(N)$}
\label{R00cal}

% QKIIIEqNummer
We show how the expressions (\ref{R00}) are derived from Eq.~(141)
in Ref.~\cite{QKIII}. 

\subsection{Simplifying the integral}

$R_{00}$ is defined by
\end{multicols}
\Endrule
%\widetext
\begin{eqnarray}
R_{00}(N)&=&\frac{4 u^2}{(2 \pi)^5 \hbar^2} \int
d^3 {\bf u} \int d^3 {\bf K}_1 \int d^3 {\bf K}_2
\int d^3 {\bf k} \int d^3 {\bf k}' \, \delta({\bf K}_1-{\bf K}_2-{\bf k}+{\bf k}')
F({\bf K}_1,{\bf u})(1+F({\bf K}_2,{\bf u})) \nonumber \\
&& W_0({\bf u},{\bf k}) W_0({\bf u},{\bf k}')
\delta(\Delta\omega_{12}({\bf u})),
\end{eqnarray}
%\narrowtext
%\Beginrule
%\begin{multicols}{2}\noindent
where
\begin{equation}
W_0({\bf u},{\bf k})=\frac{1}{(2 \pi)^3} \int d^3{\bf v} \, \xi_{N}^* ({\bf u} + 
\frac{{\bf v}}
{2}) \xi_{N}({\bf u} - \frac{{\bf v}}{2}) \mbox{e}^{i {\bf k \cdot v}},
\end{equation}
and $\Delta \omega_{12}({\bf u})$ accounts for energy conservation.
We assume that within the range
of the condensate $F({\bf K},{\bf u})$ is constant, and that the factor $1+F({\bf K}_2,{\bf u})
\approx 1$. Integrating over ${\bf K}_2$ and defining
${\bf Q}={\bf k}-{\bf k}'$ yields
%\end{multicols}
%\Endrule
%\widetext
\begin{eqnarray}
R_{00}(N) & = & \frac{4 u^2}{(2 \pi)^5 \hbar^2} \frac{2 m}{\hbar} \int d^3 {\bf K_1} \int d^3 
{\bf Q} \, \delta({\bf Q}\cdot({\bf Q}+2{\bf K}_1))
F({\bf K}_1,0) \nonumber \\
&&\int d^3 {\bf k}' \int d^3 {\bf u} \frac{1}{(2 \pi)^6}
\int d^3 {\bf v}' \int d^3 {\bf v} \, \xi_N^*({\bf u}+\frac{\bf v}{2})
\xi_N({\bf u}-\frac{\bf v}{2})\xi_N^*({\bf u}+\frac{{\bf v}'}{2})\xi_N({\bf u}-\frac{{\bf 
v}'}{2})
\mbox{e}^{i {\bf k}'\cdot({\bf v}+{\bf v}')+i{\bf Q}\cdot {\bf v}}.
\end{eqnarray}
%\narrowtext
\Beginrule
\begin{multicols}{2}\noindent
By integrating over ${\bf k}'$ and over  ${\bf v}'$ and defining
\begin{eqnarray}
G^2({\bf Q})&=&\frac{1}{(2 \pi)^3} \int d^3 {\bf u} \int d^3 {\bf v} \, \left| \xi_N({\bf 
u}+\frac{\bf
v}{2})\right|^2 \nonumber \\
&& \left| \xi_N({\bf u}-\frac{\bf v}{2})\right|^2 \mbox{e}^{i {\bf Q}\cdot{\bf v}}
\end{eqnarray}
we obtain
\begin{eqnarray}
R_{00}(N)&=&\frac{4 u^2}{(2 \pi)^5 \hbar^2} \frac{2 m}{\hbar} \int d^3 {\bf K}_1 \int d^3 
{\bf Q} \, \delta({\bf Q}\cdot({\bf Q}+2{\bf K}_1)) \nonumber \\
&& F({\bf K}_1,0) G^2({\bf Q}).
\end{eqnarray}
Now we approximate ${\bf Q}+2{\bf K}_1 \approx 2{\bf K}_1$, since $G^2({\bf Q})$ is 
sharply peaked around ${\bf Q}={\bf 0}$ compared to the width of $F({\bf K}_1,0)$. 
Replacing the notation ${\bf K}_1$ by ${\bf K}$, choosing the $K_z$ axis
in the direction of ${\bf Q}$, we obtain
\begin{eqnarray}
R_{00}(N)&=&\frac{4 u^2 m}{(2 \pi)^5 \hbar^3} \int d^3 {\bf Q} \int dK_x \int dK_y
\int dK_z \delta(QK_z) \nonumber \\
&& \mbox{e}^{(-\frac{\hbar^2(K_x^2+K_y^2+K_z^2)}{2m} + \mu)/kT} G^2({\bf Q}).
\end{eqnarray}
Here $Q$ denotes the modulus of ${\bf Q}$. Then we do the integration over ${\bf K}$ to obtain
\begin{equation}
R_{00}(N)=\frac{8 u^2 \pi m^2 k T}{(2 \pi)^5 \hbar^5} \mbox{e}^{\mu/kT} \int d^3 {\bf Q}
\, \frac{G^2({\bf Q})}{Q}.
\end{equation}
So far we have not assumed a particular form of the trapping potential or the condensate
wave function. These properties are contained in the function $G({\bf Q})$.

\subsection{Calculating the function $G({\bf Q})$ for the harmonic oscillator}

The trapping potential is of the form $V_T({\bf x})=a x^2+b y^2 +c z^2$ with $a=m 
\omega_x^2/2$,
$b=m \omega_y^2/2$ and $c=m \omega_z^2/2$.
We change the variables by defining ${\bf x}={\bf u}+\frac{\bf v}{2}$ and 
${\bf y}={\bf u}-\frac{\bf v}{2}$ and get
\begin{equation}
G^2({\bf Q})=\frac{1}{(2 \pi)^3} \int d^3 {\bf x} \int d^3 {\bf y} \, \left| \xi_N({\bf x}) 
\right|^2 \left| \xi_N({\bf y}) \right|^2 \mbox{e}^{i {\bf Q} \cdot ({\bf x}-{\bf y})},
\end{equation}
and therefore
\begin{equation}
G({\bf Q})=\frac{1}{(2 \pi)^{3/2}} \int d^3 {\bf x} \, \left| \xi_N({\bf x}) \right|^2 
\mbox{e}^{i {\bf Q} \cdot {\bf x}}.
\end{equation}
Using the Thomas-Fermi approximation for the condensate wave function,
defining $Q_x'=Q_x /\sqrt{a}$, $Q_y'=Q_y /\sqrt{b}$ and $Q_z'=Q_z /\sqrt{c}$, and changing
the variables $x'=x \sqrt{a}$, $y'=y \sqrt{b}$ and $z'=z \sqrt{c}$, simplifies
the integral to 
\begin{equation}
G({\bf Q})=\frac{1}{(2 \pi)^{3/2} \sqrt{a b c} N u} \int_{{\bf x}^{'^2}<\mu_N} d^3 {\bf x}' 
(\mu_N - {\bf x}^{'2}) \mbox{e}^{i {\bf Q}' \cdot {\bf x}'}.
\end{equation}
Integrating in spherical coordinates we obtain
\begin{eqnarray}
G({\bf Q}) & = & \frac{2}{\sqrt{ 2 \pi a b c} N u Q'} \int_0^{\sqrt{\mu_N}} dr \, (r \mu_N 
-r^3) \sin
(Q'r) \nonumber \\ 
&= & \frac{4}{\sqrt{a b c} N u Q^{'5}} \left[(3- \mu_N Q'^2) 
\sin(Q'\sqrt{\mu_N}) - \right. \nonumber 
 \\
 &&   
\left. 3 \sqrt{\mu_N} Q' \cos(Q' \sqrt{\mu_N}) \right], 
\end{eqnarray}
where  $Q'=\sqrt{Q^{'2}_x+Q^{'2}_y+Q^{'2}_z}$.
\subsection{Integration over ${\bf Q}$}

Last we will now find an expression for the integral
\begin{equation}
\Omega=\int d^3 {\bf Q} \frac{G^2({\bf Q})}{Q}.
\end{equation}
We make use of the fact that $G({\bf Q})$ only depends on $Q'$ by replacing
$d^3 {\bf Q}= \sqrt{abc} \; d^3 {\bf Q}'$ using spherical coordinates, and
change the variable $Q' \rightarrow \sqrt{\mu_N} Q'=T$. The modulus $Q$ can be
expressed in terms of ${\bf Q}'$ as
$Q=Q'\sqrt{a \sin^2 \theta \cos^2 \phi +b \sin^2 \theta \sin^2 \phi +
c \cos^2 \theta}$. We obtain
\end{multicols}
\Endrule
%\widetext
\begin{equation}
\Omega=\frac{8 \mu_N^4}{\pi \sqrt{a b c} N^2 u^2} \int_0^{2 \pi} d\phi \int_0^{\pi} d\theta
\frac{\sin(\theta)}{\sqrt{a \sin^2 \theta \cos^2 \phi +b \sin^2 \theta
\sin^2 \phi + c \cos^2 \theta}} \int_0^{\infty} dT \frac{1}{T^9} \left[(3-T^2)
\sin(T) - 3 T \cos(T)\right]^2.
\end{equation}
%\narrowtext
\Beginrule
\begin{multicols}{2}\noindent
Next we integrate over $T$, define $x=\cos \theta$, 
$\alpha^2=c-a \cos^2 \phi -b \sin^2 \phi$, and
$\beta^2= (a \cos^2 \phi +b \sin^2 \phi)/\alpha^2$, and write
\begin{eqnarray}
\Omega&=&\frac{\mu_N^4}{9 \pi \sqrt{a b c} N^2 u^2} \int_0^{2 \pi} d\phi \frac{1}{\alpha} 
\int_{-1}^1 \frac{dx}{\sqrt{x^2+\beta^2}} \nonumber \\
&=& \frac{2 \mu_N^4}{9 \pi \sqrt{a b c} N^2 u^2} \int_0^{2 \pi} d \phi
\frac{\mbox{arcsinh}(\beta^{-1})}{\alpha}.
\end{eqnarray}
In the case $\alpha^2<0$ we take the modulus of $\alpha^2$ and replace $\mbox{arcsinh} 
\rightarrow
\mbox{arcsin}$. To further evaluate this integral we now assume that $a=b$.
Then $\alpha^2=c-a$ and $\beta^2=a/(c-a)$. The integration over $\phi$ then yields a factor of
$2\pi$, and we obtain the expressions given in Eqs.~(\ref{R001}) and (\ref{R002}).
$R_{00}(N)$ therefore depends on the number of particles in the condensate as $N^{-2/5}$.

\Endrule
\end{multicols}


\begin{thebibliography}{10}

%QKI
\bibitem{QKI}
{C.~W.~Gardiner} and {P.~Zoller}, Phys. Rev. A {\bf 55}, 2902 (1997).

%QKII
\bibitem{QKII}
{D.~Jaksch}, {C.~W.~Gardiner}, and {P.~Zoller}, Phys. Rev. A {\bf 56}, 575 (1997).

%QKIII
\bibitem{QKIII}
{C.~W.~Gardiner} and {P.~Zoller}, cond-mat/9712002.

\bibitem{Anglin} 
{J.~Anglin}, Phys. Rev. Lett. {\bf 79}, 6 (1997).


%Experiments
\bibitem{JILA} M.~Anderson, J.~R.~Ensher, M.~R.~Matthews,
C.~E.~Wieman, and E.~A.~Cornell,
%Observation of BEC in a dilute atomic vapor
Science {\bf 269}, 198 (1995).

\bibitem{MIT} K.~B.~Davis, M-O.~Mewes, M.~R.~Andrews. N.~J.~van Druten, D.~S.~
Durfee, D. M. Kurn, and W. Ketterle,
Phys. Rev. Lett. {\bf 75}, 3969 (1995).

\bibitem{RICE}C.~C.~Bradley, C.~A.~Sackett, J.~J.~Tollet, and  R.~Hulet,
%Evidence of BEC in an atomic gas with attractive interactions,
Phys. Rev. Lett. {\bf 75}, 1687 (1995).

\bibitem{AUSTIN}D. J.~Heinzen, see 
http://storm.ph.utexas.edu/dept/ \\ research/heinzen/bose.html

\bibitem{ROWLAND} L.~Hau, see http://amo.phy.gasou.edu/bec.html/

\bibitem{STANFORD} B.~Anderson and M.~Kasevich, see \\ http://amo.phy.gasou.edu/bec.html/

\bibitem{Rempe}
G.~Rempe, see http://amo.phy.gasou.edu/bec.html/

%Review
\bibitem{Villain} P.~Villain, M.~Lewenstein, R.~Dum, Y.~Castin, L.~
You, A.~Imamoglu, and T.~B.~A.~Kennedy, J. Mod. Optics, {\bf 44}, 1775 (1997).

\bibitem{phase1}
%Quantum phase of a BEC with an arbitrary number of atoms
{J.~Javanainen} and {S.~M.~Yoo}, Phys. Rev. Lett. {\bf 76}, 161 (1997).
%Quantum phase diffusion of a BEC
{M.~Lewenstein} and {L.~You}, {\it ibid}. {\bf 77}, 3489 (1997);
%Inhibition of coherence in trapped BEC
{A.~Imamoglu}, {M.~Lewenstein}, and {L.~You}, {\it ibid}. {\bf 78}, 2511 (1997);
{R.~Graham}, T.~Wong, M.~J.~Collet, S.~M.~Tan, and {D.~F.~Walls} Phys. Rev. A {\bf 57}, 493 (1998).

\bibitem{phase2}
%Relative phase of two BEC
{Y.~Castin} and {J.~Dalibard}, Phys. Rev. A {\bf 55}, 4330 (1997);
%Phase and phase diffusion of a Split BEC
{J.~Javanainen} and {M.~Wilkens}, Phys. Rev. Lett. {\bf 78}, 4675 (1997);
%Influence of pumping on the relative phase of twin-trap BEC
{M.~J.~Steel} and {D.~F.~Walls}, Phys. Rev. A {\bf 56}, 3832 (1997).

\bibitem{measurementofphase} For a discussion of measurement of the
phase of the bose condensate see:
%Continuous observation of interference fringes from bose condensates
J.~I.~Cirac, C.~W.~Gardiner, M.~Naraschewski, and P.~Zoller, Phys. Rev.
A 54, R3714 (1996);
%Optical measurement of the condensate phase
{A. Imamoglu} and {T.A.B. Kennedy}, {\it ibid}. {\bf 55}, R849 (1997);
%Phase dependent spectrum of scattered light from two BEC
{J. Ruostekoski} and {D.F. Walls}, {\it ibid}. {\bf 55}, 3625 (1997).
%Nondestructive optical measurement of relative phase between two BEC
{J. Ruostekoski} and {D.F. Walls}, {\it ibid}. {\bf 56}, 2996 (1997).

%Number conserving Bogoliubov method
\bibitem{NumCon}
{C.W. Gardiner}, Phys.Rev. A {\bf 56}, 1414 (1997).  

%Coherence, Correlations, and Collisions: What one learns about BEC from their decay
\bibitem{JilaCCC}
{E.A. Burt}, {R.W. Ghrist}, {C.J. Myatt}, {M.J. Holland}, {E.A. Cornell}, and {C.E. Wieman} 
Phys. Rev. Lett. {\bf 79}, 337 (1997).

%Evaporative Cooling of trapped atoms
\bibitem{Kettrev}
{W.~Ketterle} and {N.~J.~van Druten}, Adv. At. Mol. Opt. Phys. {\bf 37}, 181 (1996).

%Quantum Noise
\bibitem{QNoise} 
{C.~W.~Gardiner}, {\em Quantum Noise} (Springer Berlin 1991).

%Systematic description of the phase of a laser
\bibitem{KGLaser}
{K.~M.~Gheri}, {D.~F.~Walls}, and {M.~A.~Marte} Phys. Rev. A {\bf 46}, 6002 (1992).

\bibitem{corrinout} The correlation functions [Eqs.~(\ref{intf}) and
(\ref{cfct})] refer to  fluctuation properties inside the trap.
This must be related to the correlation functions describing 
the response of the atom counter by a proper theory of an output
coupler. In this context, see 
H.~Steck, M.~Naraschewski, and H.~Wallis, LANL preprint quant-ph/9708014;
{E.~V.~Goldstein} and {P.~Meystre}, LANL preprint physics/9710043. {R.~J.~
Ballagh}, {K.~Burnett}, and {T.~F.~Scott}, Phys. Rev. Lett. {\bf 78},
1607 (1997). {M.-O.~Mewes}, {M.~R.~Andrews}, {D.~M.~Kurn}, {D.~S.~Durfee},
{C.~G.~Townsend}, and {W.~Ketterle}, Phys. Rev. Lett. {\bf 78}, 582
(1997).

\bibitem{JilaExp}
{C.~J.~Myatt}, {E.~A.~Burt}, {R.~W.~Ghrist}, {E.~A.~Cornell}, and {C.~E.~Wieman} Phys. Rev. Lett. 
{\bf 78},
 586 (1997).

%Growth of a BEC
\bibitem{BosGro}
{C.~W.~Gardiner}, {P.~Zoller}, {R.~J.~Ballagh}, and {M.~J.~Davis},
 Phys. Rev. Lett. {bf 79}, 1793 (1997).

%Start change

%neglect quasiparticle fluctuations
\bibitem{quasinegl}
This implies that fluctuations of particles from the condensate mode to quasiparticle
modes as well as to very low-lying one-particle excitations are not included in this
approach. Calculations on fluctuations between the condensate mode and low lying excited modes 
can be found in S.~Giorgini, L.~P.~Pitaevskii and S.~Stringari, LANL preprint, cond-mat/9711065.

%QKV
\bibitem{QKV}
C.~W.~Gardiner and P.~Zoller (unpublished).

%QKVI
\bibitem{QKVI}
M.~J.~Davis, R.~J.~Ballagh, C.~W.~Gardiner and P.~Zoller (unpublished).

%Theory of an atom laser
\bibitem{AtomLaser}
M.~Holland, K.~Burnett, C.~Gardiner, J.~I.~Cirac, and P.~Zoller, Phys. Rev. A {\bf 54}, R1757 (1996). 
%End change

%Calculation of inelastic collision rates
\bibitem{Anarates1}
{P.~Fedichev}, {M.~Reynolds}, and {G.~V.~Shlyapnikov}, Phys. Rev. Lett. {\bf 77}, 2921 (1996).

%Calculation of inelastic collision rates
\bibitem{Anarates2}
{B.~D.~Esry}, {C.~H.~Greene}, {Y.~Zhou}, and {C.~D.~Lin}, J. Phys. B {\bf 29}, L51 (1996).

%Calculation of inelastic collision rates
\bibitem{Anarates3}
{H.~M.~J.~M.~Boesten}, {A.~J.~Moerdijk}, and {B.~J.~Verhaar}, Phys. Rev. A {\bf 54}, R29 (1996).

%Nonlinear effects in lasers
\bibitem{Helmutnonlin}
{H. Ritsch}, Quantum Opt. {\bf 2}, 189 (1990).

%Start change
%Neglecting fluctuations
%Bosonic stimulation in the formation of a Bose-Einstein condensate
\bibitem{negfluc}
This equation is found by neglecting all fluctuations. Inclusion of fluctuations into this
equation has not yet been done, and is clearly desirable. However, in current experiments fluctuations
in the initial conditions due to technical uncertainties between two experimental runs are much 
larger than any fluctuations one could expect from a fundamental point of view. See
H.-J.~Miesner, D.~M.~Stamper-Kurn, M.~R.~Andrews, D.~S.~Durfee, S.~Inouye, and W.~ Ketterle, Science
{\bf 279}, 1005 (1998).

%Role of quasiparticles in the growth of a trapped Bose-Einstein condensate
\bibitem{roleofquasi}
C.~W.~Gardiner, M.~D.~Lee, R.~J.~Ballagh, M.~J.~Davis, and P.~Zoller, LANL preprint, cond-mat/9801027.
%End change

\bibitem{R00just}
The use of this approximation for the terms involving $R_{00}$ leads to the decay
rate of $16/25 \bar{R}_{00}$ instead of $\bar{R}_{00}$, as expected from the expression
for $\tau_{\bar{N}}$. However, for parameters chosen according to current experimental
setups, the exact value of the rate $\bar{R}_{00}$ has only a minor influence on the 
correlation function $\langle B^{\dagger}(t) B(s) \rangle$. 

%Quantum optics
\bibitem{Kerr}
{D.~F.~Walls} and {G.~J.~Milburn} {\em Quantum Optics}, Springer (1994). 

% Systems driven by colored squeezed noise, the atomic absorption spectrum
\bibitem{colornoise}
{H. Ritsch} and {P. Zoller}, Phys. Rev. A {\bf 38}, 4657 (1988). 
    
%Dynamics and measurement of the absolute phase in macroscopic quantum systems
\bibitem{Sols} 
{F.~Sols} and {R.~A.~Hegstrom}, {\em Fundamental Problems in Quantum Physics}, 
(Kluwer Dodrecht, 1995), p.~299.

%Start change
%Collective and single-particle excitations of a trapped Bose gas
\bibitem{Stringari2}
F.~Dalfovo, S.~Giorgini, M.~Guilleumas, L.~Pitaevskii, and S.~Stringari, Phys. Rev. A {\bf 56}, 3840 (1997).

%Conserving and Gapless Approximations for an Inhomogeneous Bose Gas at Finite Temperatures
\bibitem{HFapprox}
A.~Griffin, Phys. Rev. B {\bf 53}, 9341 (1996).

%Finite temperature excitations of a trapped Bose Gas
\bibitem{Griffin}
D.~A.~W.~Hutchinson, E.~Zaremba, and A.~Griffin, Phys. Rev. Lett. {\bf 78}, 1842 (1997).
%End change

\end{thebibliography}
\end{document}